\documentclass[aps,prl,twocolumn,longbibliography,floatfix,superscriptaddress]{revtex4-2}

\usepackage{amsmath,amssymb}
\usepackage{graphicx,braket}
\usepackage[colorlinks=true,linkcolor=blue,citecolor=blue,urlcolor=blue]{hyperref}
\usepackage{booktabs}
\usepackage[normalem]{ulem}
\usepackage{xcolor,comment}
\usepackage{color}
\usepackage{balance}
\makeatletter
\def\@bibitem#1{%
  \ifnum\value{enumiv}=35 \balance \fi 
  \item
}
\makeatother

\usepackage[T1]{fontenc}
\usepackage[utf8]{inputenc}
\usepackage{newtxtext}

\usepackage{amsmath}
\usepackage{mathtools}
\usepackage{amsfonts}
\usepackage{bm}
\usepackage{amssymb}

\usepackage{newtxmath}

\begin{document}
\title{Long-range resonances in quasiperiodic many-body localization}
\author{Ashirbad Padhan}
\email{ashirbad.padhan@irsamc.ups-tlse.fr}
\affiliation{Univ. Toulouse, CNRS, Laboratoire de Physique Th\'eorique, Toulouse, France}
\author{Jeanne Colbois}
\email{jeanne.colbois@cnrs.fr}
\affiliation{Institut N\'eel, CNRS \& Universit\'e Grenoble Alpes, 38000 Grenoble, France}
\author{Fabien Alet}
\email{fabien.alet@cnrs.fr}
\affiliation{Univ. Toulouse, CNRS, Laboratoire de Physique Th\'eorique, Toulouse, France}
\author{Nicolas Laflorencie}
\email{nicolas.laflorencie@cnrs.fr}
\affiliation{Univ. Toulouse, CNRS, Laboratoire de Physique Th\'eorique, Toulouse, France}

\date{\today}

\begin{abstract}
We investigate long-range resonances in quasiperiodic many-body localized (MBL) systems. Focusing on the Heisenberg chain in a deterministic Aubry-Andr\'{e}
 potential, we complement standard diagnostics by analyzing the structure of long-distance pairwise correlations at high energy. Contrary to the expectation that the ergodic-MBL transition in quasiperiodic systems should be sharper due to the absence of Griffiths regions, we uncover a broad unconventional regime at strong quasiperiodic potential, characterized by fat-tailed distributions of longitudinal correlations at long distance. This reveals the presence of atypical eigenstates with strong long-range correlations in a regime where standard diagnostics indicate stable MBL. We further identify these anomalous eigenstates as quasi-degenerate pairs of resonant cat states, which exhibit entanglement at long distance. These findings advance the understanding of quasiperiodic MBL and identify density-correlation measurements in ultracold atomic systems as a probe of long-range resonances.

\end{abstract}

\maketitle

{\textit{Introduction---}}
The intricate interplay between disorder and interactions in quantum many-body systems has emerged as a central theme in condensed matter and atomic physics, leading to the discovery of many-body localization (MBL)~\cite{Fleishman_1980,jacquod_emergence_1997,Altshuler_1997, gornyi_interacting_2005, Basko_2006, Oganesyan_2007, Pal_2010}. Unlike conventional ergodic systems, which are well-described by the eigenstate thermalization hypothesis (ETH)~\cite{deutsch_quantum_1991, srednicki_chaos_1994}, MBL systems fail to thermalize, retaining memory of their initial state for infinitely long times~\cite{nandkishore_many-body_2015, alet_many-body_2018, abanin_colloquium_2019, sierant_many-body_2025}. This phenomenon has been extensively studied in disordered one-dimensional spin chains, such as the random-field XXZ chain~\cite{znidaric_many-body_2008,luitz_many-body_2015, sierant_polynomially_2020, colbois_interaction_2024}. Despite remarkable progress, there is still no consensus regarding the stability of the MBL phase at the thermodynamic limit~\cite{suntajs_quantum_2020,suntajs_ergodicity_2020,sels_dynamical_2021,sirker_particle_2022,weisse_operator_2025,sierant_thouless_2020,panda_can_2020,abanin_distinguishing_2021,sierant_challenges_2022,morningstar_avalanches_2022}. So far, two potential instabilities of MBL that could drive the transition have been discussed: thermalizing avalanches~\cite{de_roeck_stability_2017,thiery_many-body_2018} triggered by rare, locally ergodic Griffiths regions (inevitable in infinite systems), as studied using toy models in finite-size numerics~\cite{luitz_how_2017,suntajs_ergodicity_2022,peacock_many-body_2023,ha_many-body_2023,colmenarez_ergodic_2024,szoldra_catching_2024,pawlik_many-body_2024,pawlik2025unconventionalthermalizationlocalizedchain}, and the proliferation of many-body resonances~\cite{gopalakrishnan_low_2015,kjall_many-body_2018,villalonga_eigenstates_2020,garratt_local_2021,garratt_resonant_2022,Tikhonov_Eigenstate_2021,Tikhonov_From_2021,morningstar_avalanches_2022,crowley_constructive_2022,ha_many-body_2023,biroli_large-deviation_2024,scoquart_role_2024,colbois_statistics_2024,laflorencie_cat_2025}, ultimately leading to delocalization.

While much of the research has focused on random disorder, quasiperiodic (QP) potentials have emerged as a deterministic alternative for inducing localization~\cite{harper_single_1955, AubryAndre,Iyer_many-body_2013, Naldesi_2016, Setiawan_2017, lee_many_2017, Khemani_2017, Bera_2017, nag_many_2017, bar_lev_transport_2017, Setiawan_2017, _nidari__2018, Weidinger_2018,Zhang_2018,Doggen_2019, Xu_2019, Weiner_2019, Sierant_2019, Aramthottil_2021, Singh_2021_qp, Agrawal_2022, _trkalj_2022, Vu_2022, sierant_challenges_2022, Tu_2023, Thomson_2023, Prelov_ek_2023, Falc_o_2024, Prasad_2024,ghosh_scaling_2025}. These systems, which can be experimentally realized with ultracold atoms in bichromatic optical lattices~\cite{roati_anderson_2008,Schreiber_2015, bordia_probing_2017,L_schen_2017,Lukin_2019, Rispoli_2019, L_onard_2023,hur2025stabilitymanybodylocalizationdimensions} or superconducting qubits arrays~\cite{li_observation_2023}, lack the rare, weak-disorder regions that characterize random systems.
This absence of Griffiths regions led to the early suggestion~\cite{Iyer_many-body_2013} that the MBL phase diagram 
could be different for QP and random potentials, as 
further strengthened by the avalanche scenario~\cite{Agrawal_2022,Tu_avalanche_2023}. Correspondingly, the asymptotic MBL-ergodic transitions in QP and random systems have been argued to belong to different universality classes ~\cite{Khemani_2017,Zhang_2018,Agrawal_2020,sun_characterizing_2025}, with evidence coming from different sample-to-sample fluctuations in finite-size numerics~\cite{Khemani_2017,ghosh_scaling_2025}. Phenomenologically, the MBL phase appears more robust in QP systems, both in numerics~\cite{Aramthottil_2021,Doggen_2019}, where the  MBL-ergodic crossover has been characterized as sharper~\cite{Khemani_2017,Falc_o_2024}, and also in ultracold atoms experiments in two dimensional finite systems~\cite{hur2025stabilitymanybodylocalizationdimensions}. However, some dynamical aspects look very similar as one approaches the transition in both QP and random systems~\cite{Weiner_2019}. Further, no systematic analysis seeking many-body resonances or long-range correlations at strong disorder has been performed in QP systems. This gap is particularly striking given their potential impact on the MBL phase on observable sizes~\cite{Tikhonov_Eigenstate_2021,morningstar_avalanches_2022,colbois_statistics_2024,giri_from_2025,biroli_large-deviation_2024, miranda_large-deviation_2025}.

In this Letter, we remedy this situation by considering the statistics of long-distance correlations in a QP Heisenberg spin chain, where, instead of {\it{rare regions}}, we uncover the presence of {\it{rare eigenstates}} that play a crucial role.
Our analysis reveals a striking similarity to the instabilities reported in random systems~\cite{colbois_statistics_2024}, with the unexpected appearance of a broad, fat-tailed distribution of large spin correlations at long distances, even in the case of a deterministic quasi-disorder.
These “rare events” take the form of anomalous eigenstates made of quasi-degenerate pairs of resonant cat states~\cite{laflorencie_cat_2025}, characterized by strong ${\cal{O}}(1)$ long-distance correlation and entanglement. 
Such atypical states emerge in a regime where standard diagnostics (e.g., spectral statistics, entanglement entropy, multifractality) already exhibit well-converged MBL behavior. 
Their proliferation indicates clear instability when approaching the critical regime from the MBL side, as evidenced from the rapid growth of the longitudinal correlation length with system size.

Our results for a deterministic QP model highlight the importance of examining the full distribution of long-distance correlations to capture the influence of rare resonant eigenpairs across the ergodic–MBL transition. We further demonstrate that instabilities driven by anomalous eigenstates are not unique to random disorder, pointing to the possible existence of a universal mechanism.

\begin{figure*}[t!]
    \includegraphics[width=2\columnwidth]{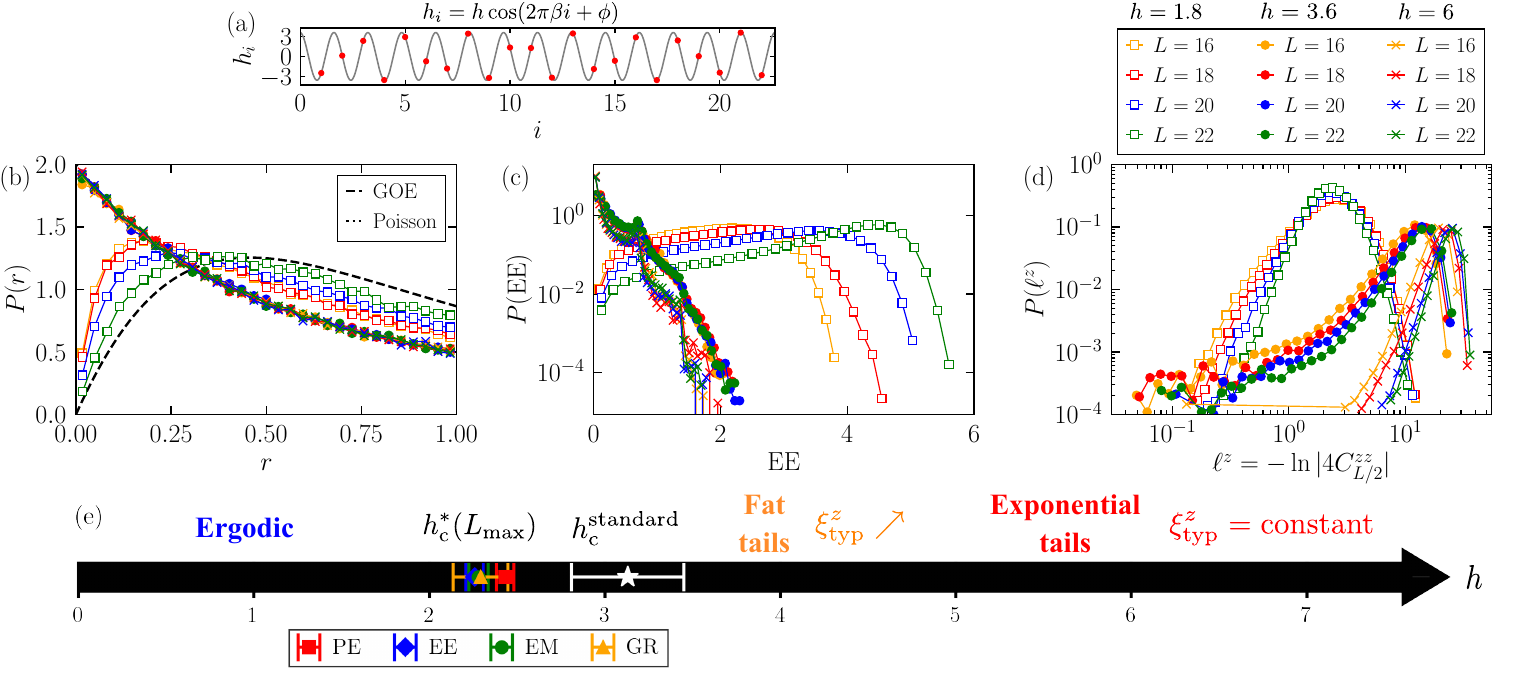}
    \caption{
Overview of the results for the 
QP Heisenberg spin chain model (Eq.~\eqref{eq:H} with $\Delta=1$). 
(a) QP field profile $h_i$ for $h=3.6$ and $\phi \approx 0.0554$. The red dots mark lattice sites $i$ for $L=22$. 
(b) Probability distribution of the gap ratio $P(r)$ for system sizes $L=16,18,20,22$ at field strengths $h=1.8$, $3.6$, and $6$. 
For $h=1.8$, the data approach GOE statistics with increasing $L$, while for $h=3.6$ and $h=6$, they follow Poisson statistics, signaling MBL. 
(c) Corresponding distributions of the half-chain entanglement entropy $P(\mathrm{EE})$ for the same parameters. 
$P(\mathrm{EE})$ shows volume-law behavior for $h=1.8$ and shifts toward localized behavior for $h=3.6$ and $h=6$. 
(d) Distribution of the rescaled longitudinal correlations $P(\ell^z)$, where $\ell^z=-\ln|4C^{zz}_{L/2}|$. 
For $h=1.8$, $P(\ell^z)$ is characteristic of the ergodic phase. 
At $h=3.6$, it develops pronounced fat tails that decay slowly with $L$, revealing eigenstates with strong long-range correlations. 
For $h=6$, $P(\ell^z)$ exhibits an exponential tail, consistent with a fully localized regime. 
(e) Schematic phase diagram {\emph{vs.}} $h$, showing the ergodic phase, a fat-tail regime where the typical longitudinal correlation length $\xi_{\mathrm{typ}}^z$ increases with system size, and an exponential-tail regime where $\xi_{\mathrm{typ}}^z$ remains constant. 
The finite-size critical region $h_{\rm c}^*(L_{\max})$, obtained from the crossings between the largest available system sizes ($L=20$ and $22$), is indicated by the colored symbols, while the extrapolated critical point $h_{\rm c}^{\mathrm{standard}}=3.13(32)$, obtained from all the standard observables, is marked by the white star (see End Matter).
\label{fig:fig1}}
\end{figure*}

{\textit{Overview of the main results---}}
We consider the QP spin-$1/2$ XXZ chain, described by the Hamiltonian
\begin{equation}
\mathcal{H} = \sum_{i=1}^{L} \left( S_i^x S_{i+1}^x + S_i^y S_{i+1}^y + \Delta S_i^z S_{i+1}^z + h_i S_i^z \right),
\label{eq:H}
\end{equation}
where $S_i^\alpha$ ($\alpha = x, y, z$) are spin-$1/2$ operators on site $i$, with periodic boundary conditions.
The Ising anisotropy $\Delta$ controls the interaction strength, and the longitudinal field follows the QP Aubry-Andr\'e~\cite{AubryAndre} form $h_i = h \cos(2\pi \beta i + \phi)$
with $\beta = (\sqrt{5} - 1)/2$ the inverse golden ratio, $h$ the modulation strength, and $\phi \in [0, 2\pi)$ a global phase shift uniformly sampled for ensemble averaging. 
We restrict the discussion to the Heisenberg case $\Delta=1$ for which we analyze
key observables for a very large number of disorder samples and eigenstates (see End Matter). 

While conventional indicators (see End Matter) such as the gap ratio (GR), participation entropy (PE), extreme magnetization (EM), and entanglement entropy (EE) capture the onset of localization, they are less sensitive to rare events and may miss subtle, system-spanning fluctuations associated with rare resonant states~\cite{morningstar_avalanches_2022,ha_many-body_2023}. 
To go beyond these standard diagnostics, we focus on the statistics of spin correlations at half-chain~\footnote{Due to the Hamiltonian commuting with $S^z_{\rm{total}}$, $x$ and $y$ components are equivalent by symmetry (with $\langle S_i^x\rangle=0$), while $\langle S_i^z\rangle\neq 0$ for longitudinal correlations.}
\begin{equation}
C_{i,i+L/2}^{\alpha\alpha} = \langle S_i^{\alpha} S_{i+L/2}^{\alpha} \rangle - \langle S_i^{\alpha} \rangle \langle S_{i+L/2}^{\alpha} \rangle, \quad \alpha=x,z 
\label{eq:corr}
\end{equation}
which capture the real-space structure of eigenstates, and are therefore more sensitive to rare resonances involving distant sites~\cite{colbois_interaction_2024,colbois_statistics_2024,laflorencie_cat_2025}. Furthermore, they are directly accessible to experiments~\cite{endres_observation_2011,Lukin_2019,Rispoli_2019,L_onard_2023}.

Fig.~\ref{fig:fig1} displays the distributions of three key quantities: GR, half-chain EE, and the longitudinal mid-chain correlation function, rescaled as $\ell^z = -\ln|4C^{zz}_{L/2}|$, all computed in mid-spectrum eigenstates for three representative QP field strengths $h=1.8,\,3.6,\,6$, and system sizes~{$L = 16-22$}. For simplicity, we denote $C_{i,i+L/2}^{zz}$ as $C_{L/2}^{zz}$.

In Fig.~\ref{fig:fig1}(b), the GR distribution $P(r)$ unambiguously approaches GOE statistics with increasing $L$ for $h=1.8$, while for $h=3.6$ and $h=6$, it exhibits  
Poisson statistics, a hallmark of MBL physics~\cite{Oganesyan_2007,Pal_2010}. Similarly, the distributions of half-chain EE shown in Fig.~\ref{fig:fig1}(c) reflect the same qualitative differences. For $h=1.8$ the peak of $P(\text{EE})$ shifts with increasing system size, consistent with volume-law behavior, while for $h=3.6$ and $h=6$, $P(\text{EE})$ is independent of the size, shows a peak at zero, characteristic of MBL, along with a secondary peak at the maximal 2-site entanglement, $\text{EE}=\ln 2$, {generally attributed~\cite{bauer_area_2013}} to short-distance entanglement across the cut.

In contrast with standard observables which show no qualitative difference between $h = 3.6$ and $h = 6$, the distribution of long-range correlations $P(\ell^z)$ displays a striking distinction [Fig.~\ref{fig:fig1}(d)]. For $h = 1.8$, $P(\ell^z)$ is characteristic of the ergodic phase~\cite{colbois_interaction_2024}: the distribution narrows and its peak sharpens with increasing $L$. For $h = 6$, $P(\ell^z)$ exhibits strongly suppressed (presumably exponential) tails at small $\ell^z$ (large $C^{zz}_{L/2}$), decaying rapidly with $L$. This behavior, reminiscent of Anderson localization~\cite{SM}, is consistent with a fully developed MBL phase. At intermediate strength $h = 3.6$, however, $P(\ell^z)$ displays slowly decaying fat tails that persist with increasing system size, signaling the presence of rare eigenstates with anomalously strong long-range correlations. The stark contrast between $h = 3.6$ and $h = 6$ provides compelling evidence for an instability that usual observables fail to capture. The schematic phase diagram in Fig.~\ref{fig:fig1}(e) summarizes this behavior: an ergodic-MBL transition is estimated at $h_{\mathrm{c}}^{\text{standard}}\approx 3$ based on standard diagnostics (see also below), followed by a crossover regime where $P(\ell^z)$ evolves from rare-event-dominated, fat-tailed distributions to exponentially suppressed forms. {We emphasize that $h_{\mathrm{c}}^{\text{standard}}\approx 3$ is obtained after extrapolation to the thermodynamic limit (see App.~\ref{sec:fss}), and that on the system sizes considered in Fig.~\ref{fig:fig1} all standard observables display for $h=3.6$ a clear MBL behavior.}

\begin{figure}[t!]
    \centering
    \includegraphics[width=\columnwidth]{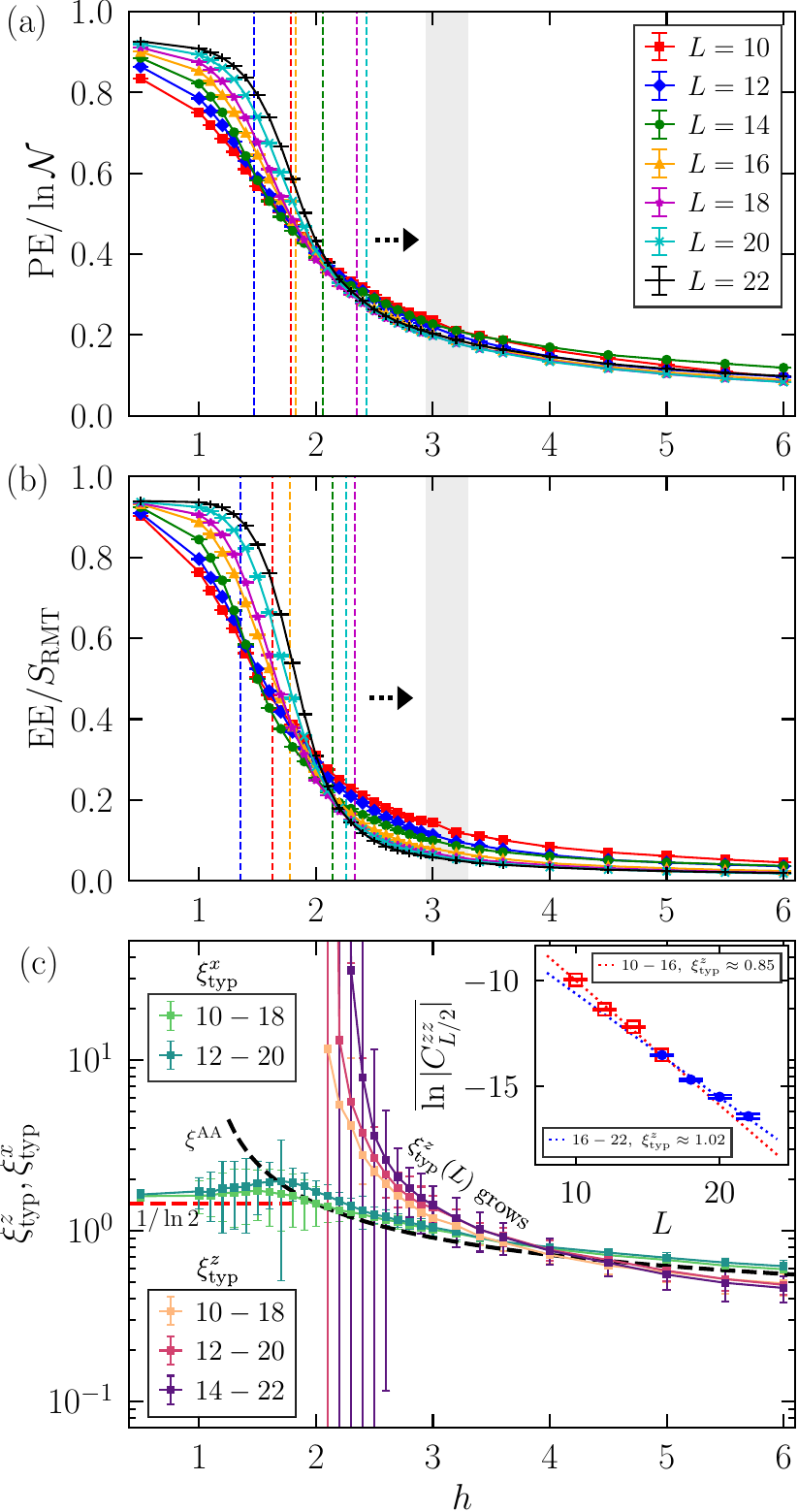}
    \caption{Scaled (a) participation and (b) half-chain entanglement entropies, 
and 
(c) typical longitudinal ($\xi_{\mathrm{typ}}^z$) and transverse ($\xi_{\mathrm{typ}}^x$) correlation lengths, all plotted against the QP field strength $h$. 
The dashed lines in (a,b) indicate the crossings between consecutive system sizes, and the shaded gray regions mark the critical regime of the ergodic–MBL transition obtained from the extrapolated finite-size crossing points (see End Matter).
In (c), the dashed line denotes the Aubry–André localization length $\xi^{\mathrm{AA}} =(\ln h)^{-1}$. Inset: $\overline{\ln |C^{zz}_{L/2}|}$ is shown {\emph{vs.}} $L$ for $h=3.6$, with dotted lines representing 4-point fits.
     \label{fig:fig2}}
\end{figure}

{\textit{Standard MBL diagnostics {\emph{vs.}} correlation lengths---}} Now we examine in more detail the differences between conventional diagnostics of the MBL transition and correlation functions at long distance. Results are presented in Fig.~\ref{fig:fig2} as a function of the QP field strength $h$. Panels (a) and (b) show the disorder- and eigenstate-averaged PE and EE, respectively, for various system sizes ($10\le L\le 22$). Both entropies are normalized to their maximal value asymptotically reached for random states (for participation: $\log {\cal N}$ where ${\cal N}$ is the size of the Hilbert space; for entanglement: $S_{\rm RMT}$, which is the RMT value~\cite{vidmar_entanglement_2017,Aramthottil_2021}) and can accurately
capture the ergodic–MBL transition. The two largest system sizes available ($L = 20$–$22$) exhibit crossings at $h \sim 2.4$, consistent with previous studies~\cite{Doggen_2019,Falc_o_2024}. As in the random-field case~\cite{sierant_polynomially_2020, colbois_interaction_2024}, we also observe a drift of the crossing points with $L$~\cite{Aramthottil_2021, Falc_o_2024}, which fitted with a $1/L$ form, eventually converges to a finite critical amplitude, $h_{\rm c} \approx 3$, in agreement with the two other observables~(see End Matter).

In order to explore possible system-wide instabilities across the ergodic-MBL transition, we complement these standard diagnostics with an analysis of mid-chain longitudinal
$\xi_{\mathrm{typ}}^z$ and transverse $\xi_{\mathrm{typ}}^x$ 
correlation lengths based on the exponential decay of the typical eigenstate- and disorder-averaged correlation function at maximal distance, see Eq.~\eqref{eq:corr}:
\begin{equation}
\ln C_{L/2,{\rm typ}}^{\alpha \alpha} := {\overline{\ln |C^{\alpha\alpha}_{L/2}|}} =:-\frac{L}{2\xi_{\rm typ}^{\alpha}}+{\cal{O}}(1).
\label{eq:Czz}
\end{equation}
Fig.~\ref{fig:fig2}(c) shows the global evolution of both longitudinal and transverse lengths as a function of $h$ and increasing system size $L$, as obtained from fits to Eq.~\eqref{eq:Czz} over sliding windows ($L=10$--$18$, $12$--$20$, $14$--$22$) to capture finite-size trends.

We first discuss the behavior of the longitudinal correlation length. In the ergodic phase, conservation of $S^{z}_{\rm total}$ results in a power-law decay of the correlator $C^{zz}_{L/2}$~\cite{Pal_2010,colbois_interaction_2024}, corresponding to $\xi_{\rm typ}^z=\infty$ formally. Conversely, deep in the MBL phase, $\xi_{\rm typ}^z$ quickly saturates to a very small finite, system-size-independent value, as observed for $h \gtrsim 5$. However, for $h< 5$ where standard quantities show a fully converged MBL response, $\xi_{\rm typ}^z$ grows noticeably with $L$, signaling enhanced correlations and a possible incipient instability of the localized phase. 
This behavior is illustrated in the inset of Fig.~\ref{fig:fig2}(c) for $h = 3.6$, where the decay of ${\overline{\ln |C^{zz}_{L/2}|}}$ evolves slowly with $L$, indicating an effective growth of $\xi_{\rm typ}^z$ with increasing system size. Larger error bars on estimates $\xi_{\rm typ}^z$ are also a marker of this instability. In the absence of a conservation law in the $x$ direction, the transverse correlation length $\xi_{\rm typ}^x$ has a very different behavior and remains independent of $L$ across the entire range. At large $h$, it follows the noninteracting Aubry-André localization length, $\xi^{\rm AA}=(\ln{h})^{-1}$, and then converges towards  $(\ln 2)^{-1}$ in the ergodic phase, consistent with the random-state prediction~\cite{colbois_statistics_2024}.
Similarly to the random-field case, $\xi_{\rm typ}^x$ is not sensitive to the MBL transition, in stark contrast with the $zz$ response.

{\textit{Cat states---}}  The fat tail observed in the distribution of mid-chain correlations $P(\ell^z)$, and the associated anomalous growth of $\xi_{\rm typ}^z$ with $L$, call for a closer examination of the eigenstate structure in the regime $h\sim 3-5$. Guided by recent findings in the random-field case~\cite{laflorencie_cat_2025}, we explore whether the strong enhancement of the longitudinal response at long distances in the QP case could result from an unusual family of eigenstates forming resonant cat-like states. A minimal  description of these cat pairs can be captured by the ansatz:

\begin{eqnarray}
    {\ket {\Phi_{\pm}^{\rm cats}}}&\sim & \Bigl({\ket{ \uparrow_{i}\downarrow_j\downarrow_{k}\uparrow_{\ell}}}
    \pm {\ket{\downarrow_{i}\uparrow_j\uparrow_{k}\downarrow_{\ell}}}\Bigr)\otimes {\ket{\psi_{L-4}}},
    \label{eq:ansatz}
\end{eqnarray}
where $r=i,\,j,\,k,\,\ell$ are 4 resonant sites (not necessarily neighbors) and  ${\ket{\psi_{L-4}}}$ encodes the quantum state of the $L-4$ remaining sites. Within this toy-model description of a resonant eigenpair, it is straightforward to verify that the local magnetization vanishes, $\langle S_r^z \rangle = 0$, as do the transverse correlations, $\langle S_r^{x} S_{r'}^{x} \rangle = \langle S_r^{y} S_{r'}^{y} \rangle = 0$ ($r \neq r'$). In contrast, the longitudinal response is maximal $\langle S_r^{z} S_{r'}^{z} \rangle - \langle S_r^{z} \rangle \langle S_{r'}^{z} \rangle = \pm 1/4$, with a sign depending on the relative orientation within the pair $(r, r')$.

\begin{figure}[t!]
    \centering
    \includegraphics[width=\columnwidth]{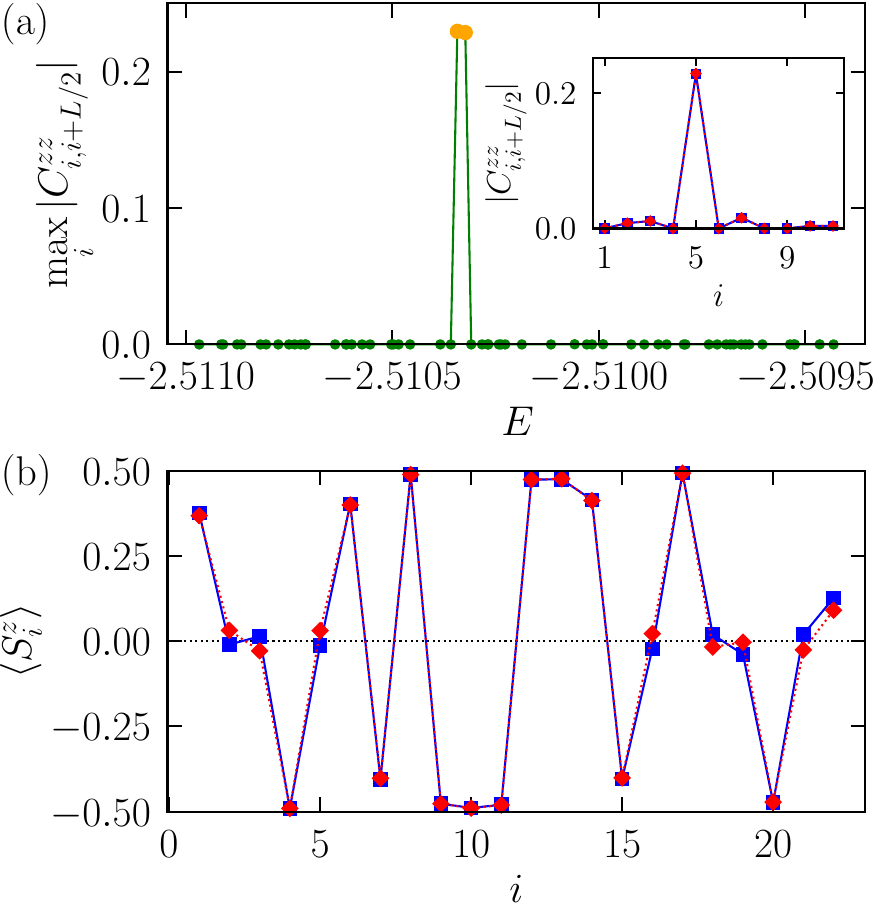}
    \caption{(a) Maximum longitudinal correlation $\max\limits_i |C_{i,i+L/2}^{zz}|$ {\emph{vs.}} energy $E$ near the middle of the spectrum for $L=22$, $h=3.6$, and $\phi \approx 0.0554$ (corresponding to Fig.~\ref{fig:fig1}(a)).  Two eigenstates with exceptionally large central correlations are highlighted by orange circles. The inset shows the real-space correlations $|C_{i,i+L/2}^{zz}|$ for these two marked states. 
(b) Local magnetization profiles $\langle S_i^z \rangle$ for the same nearly degenerate eigenstates. The blue squares correspond to the state with $|C_{L/2}^{zz}| \approx 0.2286$, and the red diamonds to the state with $|C_{L/2}^{zz}| \approx 0.2295$.\label{fig:cats}}
\end{figure}

The validity of this ansatz Eq.~\eqref{eq:ansatz} can be checked against exact numerics in the intermediate regime. In Fig.~\ref{fig:cats} we show results for a particular sample showing large correlations, governed by the QP Hamiltonian Eq.~\eqref{eq:H} with $L=22$ sites, a potential strength $h=3.6$, and a random phase $\phi = 0.055410699678\ldots$. In panel (a) we first show the mid-chain correlators $|C_{L/2}^{zz}|$ as a function of the total energy $E$ near the middle of the spectrum. While most eigenstates in this regime exhibit exponentially small correlations, we observe two nearly-degenerate eigenstates (orange circles) that both feature anomalously large correlations, $C_{L/2}^{zz} \approx 0.23$. Their energies $E^{\pm} = -2.51033288\ldots~\pm {\Delta E}/{2}$ present a tiny energy splitting, $\Delta E \approx 1.922 \times 10^{-5}$, a value comparable with the average many-body level spacing $\sim 1.5\times 10^{-5}$.

In real space, these two eigenstates have a remarkably similar structure. Both display equally strong longitudinal correlations at maximal distance between the same sites $(5,16)$. Moreover, their local magnetization profiles $\langle S_i^z \rangle$, shown in Fig.~\ref{fig:cats}(b), reveal striking patterns. In both eigenstates, the profiles are virtually identical, with several spins remaining almost fully polarized $\langle S_i^z\rangle \simeq \pm 1/2$, while the others fluctuate ($\langle S_i^z\rangle \simeq 0$),
like for instance the nearest neighbor pairs $(2-3)$ and $(18-19)$, but most importantly the strongly correlated spins at maximal distance $(5-16)$. As a result, this particular pair of eigenstates is remarkably well captured by the ansatz state in Eq.~\eqref{eq:ansatz}, with eight fluctuating sites and fourteen frozen spins (corresponding to $\ket{\psi_{L-8}}$ being a product state in the $\{S^z\}$ basis), which confirms the existence of long-range \emph{cat states}, as a coherent superposition of macroscopically distinct spin configurations.

Building on this concrete example and the corresponding results in the random-field model~\cite{laflorencie_cat_2025},  we expect that the fat-tailed instability in the distribution of correlations, and the effective growth of $\xi_{\rm typ}^z(L)$ observed within the range of $h$ previously identified as MBL, are direct consequences of such long-range-correlated cat states.
Our results for a deterministic QP potential also indicate that rare Griffiths regions (of anomalously low disorder) are not required to stabilize such anomalous eigenstates. To the best of our knowledge, this is the first time such long-range many-body resonances are directly reported for the QP potential.
A full exhaustive exploration of resonant eigenpairs and their scaling with system size or QP potential strength is left for future studies.

{\textit{Conclusions---}}
In this work, we have presented evidence for rare atypical eigenstates in the MBL regime  near the ergodic transition of deterministic QP chains, described by resonant cat-state pairs.
By analyzing longitudinal spin–spin correlation functions, we numerically demonstrated that they display fat-tailed distributions and a pronounced growth of the longitudinal correlation length $\xi_{\rm typ}^{z}(L)$ within a regime previously identified as localized by standard MBL diagnostics. This observation, made in a model devoid of statistical fluctuations inherent to random disorder, suggests that the observed instability does not originate from rare regions or ergodic inclusions. This questions the view that the transition is sharper~\cite{Khemani_2017,Aramthottil_2021,Falc_o_2024} in {\it finite-size} QP systems. Rather, our results suggest a scenario in which the stability of MBL phase, as observed on finite-size systems available to simulations and experiments, is generically threatened upon lowering the disorder strength by rare, long-range many-body resonances, which are detectable using the set of finite-size diagnostics that we use here. This may explain the observed instability of local integrals of motion at relatively strong disorder reported in Ref.~\cite{Singh_2021_qp}. At the experimental level, density correlations can now be accessed directly~\cite{endres_observation_2011,Lukin_2019,Rispoli_2019,L_onard_2023,Tarik_correlations_2025}, making our results potentially testable in the laboratory. Another experimentally-accessible dynamical signature of these rare long-range entangled states could be ``creep'' dynamics~\cite{Weiner_2019} observed in QP and random systems.\\

{\textit{Acknowledgments---}}
We gratefully acknowledge I. Bloch, M. Filippone, K. Pawlik and J. Zakrzewski for inspiring discussions and collaborations on related projects. This work has been partly supported by  HQI initiative (www.hqi.fr) and is supported by France 2030 under the French National Research Agency award number ANR- 22-PNCQ-0002. This work is also supported by the ANR research grant ManyBodyNet No. ANR-24-
CE30- 5851, and benefited from the support of the Fondation Simone et Cino Del Duca.
We acknowledge the use of HPC resources from CALMIP (grants 2024-P0677 and 2025-P0677) and GENCI (projects A0150500225 and A0170500225). We also acknowledge the use of the PETSc~\cite{petsc-user-ref,petsc-efficient}, SLEPc~\cite{slepc-toms,Hernandez:2005:SSF,slepc-manual}, MUMPS~\cite{MUMPS1,MUMPS2}, and Strumpack~\cite{Strumpack} sparse linear algebra libraries.\\

\bibliography{references,mbl}

\begin{thebibliography}{115}%
\makeatletter
\providecommand \@ifxundefined [1]{%
 \@ifx{#1\undefined}
}%
\providecommand \@ifnum [1]{%
 \ifnum #1\expandafter \@firstoftwo
 \else \expandafter \@secondoftwo
 \fi
}%
\providecommand \@ifx [1]{%
 \ifx #1\expandafter \@firstoftwo
 \else \expandafter \@secondoftwo
 \fi
}%
\providecommand \natexlab [1]{#1}%
\providecommand \enquote  [1]{``#1''}%
\providecommand \bibnamefont  [1]{#1}%
\providecommand \bibfnamefont [1]{#1}%
\providecommand \citenamefont [1]{#1}%
\providecommand \href@noop [0]{\@secondoftwo}%
\providecommand \href [0]{\begingroup \@sanitize@url \@href}%
\providecommand \@href[1]{\@@startlink{#1}\@@href}%
\providecommand \@@href[1]{\endgroup#1\@@endlink}%
\providecommand \@sanitize@url [0]{\catcode `\\12\catcode `\$12\catcode `\&12\catcode `\#12\catcode `\^12\catcode `\_12\catcode `\%12\relax}%
\providecommand \@@startlink[1]{}%
\providecommand \@@endlink[0]{}%
\providecommand \url  [0]{\begingroup\@sanitize@url \@url }%
\providecommand \@url [1]{\endgroup\@href {#1}{\urlprefix }}%
\providecommand \urlprefix  [0]{URL }%
\providecommand \Eprint [0]{\href }%
\providecommand \doibase [0]{https://doi.org/}%
\providecommand \selectlanguage [0]{\@gobble}%
\providecommand \bibinfo  [0]{\@secondoftwo}%
\providecommand \bibfield  [0]{\@secondoftwo}%
\providecommand \translation [1]{[#1]}%
\providecommand \BibitemOpen [0]{}%
\providecommand \bibitemStop [0]{}%
\providecommand \bibitemNoStop [0]{.\EOS\space}%
\providecommand \EOS [0]{\spacefactor3000\relax}%
\providecommand \BibitemShut  [1]{\csname bibitem#1\endcsname}%
\let\auto@bib@innerbib\@empty
\bibitem [{\citenamefont {Fleishman}\ and\ \citenamefont {Anderson}(1980)}]{Fleishman_1980}%
  \BibitemOpen
  \bibfield  {author} {\bibinfo {author} {\bibfnamefont {L.}~\bibnamefont {Fleishman}}\ and\ \bibinfo {author} {\bibfnamefont {P.~W.}\ \bibnamefont {Anderson}},\ }\bibfield  {title} {\bibinfo {title} {Interactions and the anderson transition},\ }\href {https://doi.org/10.1103/PhysRevB.21.2366} {\bibfield  {journal} {\bibinfo  {journal} {Phys. Rev. B}\ }\textbf {\bibinfo {volume} {21}},\ \bibinfo {pages} {2366} (\bibinfo {year} {1980})}\BibitemShut {NoStop}%
\bibitem [{\citenamefont {Jacquod}\ and\ \citenamefont {Shepelyansky}(1997)}]{jacquod_emergence_1997}%
  \BibitemOpen
  \bibfield  {author} {\bibinfo {author} {\bibfnamefont {P.}~\bibnamefont {Jacquod}}\ and\ \bibinfo {author} {\bibfnamefont {D.~L.}\ \bibnamefont {Shepelyansky}},\ }\bibfield  {title} {\bibinfo {title} {Emergence of {Quantum} {Chaos} in {Finite} {Interacting} {Fermi} {Systems}},\ }\href {https://doi.org/10.1103/PhysRevLett.79.1837} {\bibfield  {journal} {\bibinfo  {journal} {Phys. Rev. Lett.}\ }\textbf {\bibinfo {volume} {79}},\ \bibinfo {pages} {1837} (\bibinfo {year} {1997})}\BibitemShut {NoStop}%
\bibitem [{\citenamefont {Altshuler}\ \emph {et~al.}(1997)\citenamefont {Altshuler}, \citenamefont {Gefen}, \citenamefont {Kamenev},\ and\ \citenamefont {Levitov}}]{Altshuler_1997}%
  \BibitemOpen
  \bibfield  {author} {\bibinfo {author} {\bibfnamefont {B.~L.}\ \bibnamefont {Altshuler}}, \bibinfo {author} {\bibfnamefont {Y.}~\bibnamefont {Gefen}}, \bibinfo {author} {\bibfnamefont {A.}~\bibnamefont {Kamenev}},\ and\ \bibinfo {author} {\bibfnamefont {L.~S.}\ \bibnamefont {Levitov}},\ }\bibfield  {title} {\bibinfo {title} {Quasiparticle lifetime in a finite system: A nonperturbative approach},\ }\href {https://doi.org/10.1103/PhysRevLett.78.2803} {\bibfield  {journal} {\bibinfo  {journal} {Phys. Rev. Lett.}\ }\textbf {\bibinfo {volume} {78}},\ \bibinfo {pages} {2803} (\bibinfo {year} {1997})}\BibitemShut {NoStop}%
\bibitem [{\citenamefont {Gornyi}\ \emph {et~al.}(2005)\citenamefont {Gornyi}, \citenamefont {Mirlin},\ and\ \citenamefont {Polyakov}}]{gornyi_interacting_2005}%
  \BibitemOpen
  \bibfield  {author} {\bibinfo {author} {\bibfnamefont {I.~V.}\ \bibnamefont {Gornyi}}, \bibinfo {author} {\bibfnamefont {A.~D.}\ \bibnamefont {Mirlin}},\ and\ \bibinfo {author} {\bibfnamefont {D.~G.}\ \bibnamefont {Polyakov}},\ }\bibfield  {title} {\bibinfo {title} {{Interacting} {Electrons} in {Disordered} {Wires}: {Anderson} {Localization} and {Low}-{T} {Transport}},\ }\href {https://doi.org/10.1103/PhysRevLett.95.206603} {\bibfield  {journal} {\bibinfo  {journal} {Phys. Rev. Lett.}\ }\textbf {\bibinfo {volume} {95}},\ \bibinfo {pages} {206603} (\bibinfo {year} {2005})}\BibitemShut {NoStop}%
\bibitem [{\citenamefont {Basko}\ \emph {et~al.}(2006)\citenamefont {Basko}, \citenamefont {Aleiner},\ and\ \citenamefont {Altshuler}}]{Basko_2006}%
  \BibitemOpen
  \bibfield  {author} {\bibinfo {author} {\bibfnamefont {D.}~\bibnamefont {Basko}}, \bibinfo {author} {\bibfnamefont {I.}~\bibnamefont {Aleiner}},\ and\ \bibinfo {author} {\bibfnamefont {B.}~\bibnamefont {Altshuler}},\ }\bibfield  {title} {\bibinfo {title} {Metal–insulator transition in a weakly interacting many-electron system with localized single-particle states},\ }\href {https://doi.org/10.1016/j.aop.2005.11.014} {\bibfield  {journal} {\bibinfo  {journal} {Annals of Physics}\ }\textbf {\bibinfo {volume} {321}},\ \bibinfo {pages} {1126} (\bibinfo {year} {2006})}\BibitemShut {NoStop}%
\bibitem [{\citenamefont {Oganesyan}\ and\ \citenamefont {Huse}(2007{\natexlab{a}})}]{Oganesyan_2007}%
  \BibitemOpen
  \bibfield  {author} {\bibinfo {author} {\bibfnamefont {V.}~\bibnamefont {Oganesyan}}\ and\ \bibinfo {author} {\bibfnamefont {D.~A.}\ \bibnamefont {Huse}},\ }\bibfield  {title} {\bibinfo {title} {Localization of interacting fermions at high temperature},\ }\href {https://doi.org/10.1103/PhysRevB.75.155111} {\bibfield  {journal} {\bibinfo  {journal} {Phys. Rev. B}\ }\textbf {\bibinfo {volume} {75}},\ \bibinfo {pages} {155111} (\bibinfo {year} {2007}{\natexlab{a}})}\BibitemShut {NoStop}%
\bibitem [{\citenamefont {Pal}\ and\ \citenamefont {Huse}(2010)}]{Pal_2010}%
  \BibitemOpen
  \bibfield  {author} {\bibinfo {author} {\bibfnamefont {A.}~\bibnamefont {Pal}}\ and\ \bibinfo {author} {\bibfnamefont {D.~A.}\ \bibnamefont {Huse}},\ }\bibfield  {title} {\bibinfo {title} {Many-body localization phase transition},\ }\href {https://doi.org/10.1103/PhysRevB.82.174411} {\bibfield  {journal} {\bibinfo  {journal} {Phys. Rev. B}\ }\textbf {\bibinfo {volume} {82}},\ \bibinfo {pages} {174411} (\bibinfo {year} {2010})}\BibitemShut {NoStop}%
\bibitem [{\citenamefont {Deutsch}(1991)}]{deutsch_quantum_1991}%
  \BibitemOpen
  \bibfield  {author} {\bibinfo {author} {\bibfnamefont {J.~M.}\ \bibnamefont {Deutsch}},\ }\bibfield  {title} {\bibinfo {title} {Quantum statistical mechanics in a closed system},\ }\href {https://doi.org/10.1103/PhysRevA.43.2046} {\bibfield  {journal} {\bibinfo  {journal} {Phys. Rev. A}\ }\textbf {\bibinfo {volume} {43}},\ \bibinfo {pages} {2046} (\bibinfo {year} {1991})}\BibitemShut {NoStop}%
\bibitem [{\citenamefont {Srednicki}(1994)}]{srednicki_chaos_1994}%
  \BibitemOpen
  \bibfield  {author} {\bibinfo {author} {\bibfnamefont {M.}~\bibnamefont {Srednicki}},\ }\bibfield  {title} {\bibinfo {title} {Chaos and quantum thermalization},\ }\href {https://doi.org/10.1103/PhysRevE.50.888} {\bibfield  {journal} {\bibinfo  {journal} {Phys. Rev. E}\ }\textbf {\bibinfo {volume} {50}},\ \bibinfo {pages} {888} (\bibinfo {year} {1994})}\BibitemShut {NoStop}%
\bibitem [{\citenamefont {Nandkishore}\ and\ \citenamefont {Huse}(2015)}]{nandkishore_many-body_2015}%
  \BibitemOpen
  \bibfield  {author} {\bibinfo {author} {\bibfnamefont {R.}~\bibnamefont {Nandkishore}}\ and\ \bibinfo {author} {\bibfnamefont {D.~A.}\ \bibnamefont {Huse}},\ }\bibfield  {title} {\bibinfo {title} {Many-{Body} {Localization} and {Thermalization} in {Quantum} {Statistical} {Mechanics}},\ }\href {https://doi.org/10.1146/annurev-conmatphys-031214-014726} {\bibfield  {journal} {\bibinfo  {journal} {Annu. Rev. Condens. Matter Phys.}\ }\textbf {\bibinfo {volume} {6}},\ \bibinfo {pages} {15} (\bibinfo {year} {2015})}\BibitemShut {NoStop}%
\bibitem [{\citenamefont {Alet}\ and\ \citenamefont {Laflorencie}(2018)}]{alet_many-body_2018}%
  \BibitemOpen
  \bibfield  {author} {\bibinfo {author} {\bibfnamefont {F.}~\bibnamefont {Alet}}\ and\ \bibinfo {author} {\bibfnamefont {N.}~\bibnamefont {Laflorencie}},\ }\bibfield  {title} {\bibinfo {title} {{Many-body localization: {An} introduction and selected topics}},\ }\href {https://doi.org/10.1016/j.crhy.2018.03.003} {\bibfield  {journal} {\bibinfo  {journal} {Comptes Rendus Physique}\ }\textbf {\bibinfo {volume} {19}},\ \bibinfo {pages} {498} (\bibinfo {year} {2018})}\BibitemShut {NoStop}%
\bibitem [{\citenamefont {Abanin}\ \emph {et~al.}(2019)\citenamefont {Abanin}, \citenamefont {Altman}, \citenamefont {Bloch},\ and\ \citenamefont {Serbyn}}]{abanin_colloquium_2019}%
  \BibitemOpen
  \bibfield  {author} {\bibinfo {author} {\bibfnamefont {D.~A.}\ \bibnamefont {Abanin}}, \bibinfo {author} {\bibfnamefont {E.}~\bibnamefont {Altman}}, \bibinfo {author} {\bibfnamefont {I.}~\bibnamefont {Bloch}},\ and\ \bibinfo {author} {\bibfnamefont {M.}~\bibnamefont {Serbyn}},\ }\bibfield  {title} {\bibinfo {title} {Colloquium: {Many}-body localization, thermalization, and entanglement},\ }\href {https://doi.org/10.1103/RevModPhys.91.021001} {\bibfield  {journal} {\bibinfo  {journal} {Rev. Mod. Phys.}\ }\textbf {\bibinfo {volume} {91}},\ \bibinfo {pages} {021001} (\bibinfo {year} {2019})}\BibitemShut {NoStop}%
\bibitem [{\citenamefont {Sierant}\ \emph {et~al.}(2025)\citenamefont {Sierant}, \citenamefont {Lewenstein}, \citenamefont {Scardicchio}, \citenamefont {Vidmar},\ and\ \citenamefont {Zakrzewski}}]{sierant_many-body_2025}%
  \BibitemOpen
  \bibfield  {author} {\bibinfo {author} {\bibfnamefont {P.}~\bibnamefont {Sierant}}, \bibinfo {author} {\bibfnamefont {M.}~\bibnamefont {Lewenstein}}, \bibinfo {author} {\bibfnamefont {A.}~\bibnamefont {Scardicchio}}, \bibinfo {author} {\bibfnamefont {L.}~\bibnamefont {Vidmar}},\ and\ \bibinfo {author} {\bibfnamefont {J.}~\bibnamefont {Zakrzewski}},\ }\bibfield  {title} {\bibinfo {title} {Many-body localization in the age of classical computing},\ }\href {https://doi.org/10.1088/1361-6633/ad9756} {\bibfield  {journal} {\bibinfo  {journal} {Reports on Progress in Physics}\ }\textbf {\bibinfo {volume} {88}},\ \bibinfo {pages} {026502} (\bibinfo {year} {2025})}\BibitemShut {NoStop}%
\bibitem [{\citenamefont {Znidaric}\ \emph {et~al.}(2008)\citenamefont {Znidaric}, \citenamefont {Prosen},\ and\ \citenamefont {Prelovsek}}]{znidaric_many-body_2008}%
  \BibitemOpen
  \bibfield  {author} {\bibinfo {author} {\bibfnamefont {M.}~\bibnamefont {Znidaric}}, \bibinfo {author} {\bibfnamefont {T.}~\bibnamefont {Prosen}},\ and\ \bibinfo {author} {\bibfnamefont {P.}~\bibnamefont {Prelovsek}},\ }\bibfield  {title} {\bibinfo {title} {Many-body localization in the {Heisenberg} {XXZ} magnet in a random field},\ }\href {http://link.aps.org/doi/10.1103/PhysRevB.77.064426} {\bibfield  {journal} {\bibinfo  {journal} {Phys. Rev. B}\ }\textbf {\bibinfo {volume} {77}},\ \bibinfo {pages} {064426} (\bibinfo {year} {2008})}\BibitemShut {NoStop}%
\bibitem [{\citenamefont {Luitz}\ \emph {et~al.}(2015)\citenamefont {Luitz}, \citenamefont {Laflorencie},\ and\ \citenamefont {Alet}}]{luitz_many-body_2015}%
  \BibitemOpen
  \bibfield  {author} {\bibinfo {author} {\bibfnamefont {D.~J.}\ \bibnamefont {Luitz}}, \bibinfo {author} {\bibfnamefont {N.}~\bibnamefont {Laflorencie}},\ and\ \bibinfo {author} {\bibfnamefont {F.}~\bibnamefont {Alet}},\ }\bibfield  {title} {\bibinfo {title} {Many-body localization edge in the random-field {Heisenberg} chain},\ }\href {https://doi.org/10.1103/PhysRevB.91.081103} {\bibfield  {journal} {\bibinfo  {journal} {Phys. Rev. B}\ }\textbf {\bibinfo {volume} {91}},\ \bibinfo {pages} {081103(R)} (\bibinfo {year} {2015})}\BibitemShut {NoStop}%
\bibitem [{\citenamefont {Sierant}\ \emph {et~al.}(2020{\natexlab{a}})\citenamefont {Sierant}, \citenamefont {Lewenstein},\ and\ \citenamefont {Zakrzewski}}]{sierant_polynomially_2020}%
  \BibitemOpen
  \bibfield  {author} {\bibinfo {author} {\bibfnamefont {P.}~\bibnamefont {Sierant}}, \bibinfo {author} {\bibfnamefont {M.}~\bibnamefont {Lewenstein}},\ and\ \bibinfo {author} {\bibfnamefont {J.}~\bibnamefont {Zakrzewski}},\ }\bibfield  {title} {\bibinfo {title} {Polynomially filtered exact diagonalization approach to many-body localization},\ }\href {https://doi.org/10.1103/PhysRevLett.125.156601} {\bibfield  {journal} {\bibinfo  {journal} {Phys. Rev. Lett.}\ }\textbf {\bibinfo {volume} {125}},\ \bibinfo {pages} {156601} (\bibinfo {year} {2020}{\natexlab{a}})}\BibitemShut {NoStop}%
\bibitem [{\citenamefont {Colbois}\ \emph {et~al.}(2024{\natexlab{a}})\citenamefont {Colbois}, \citenamefont {Alet},\ and\ \citenamefont {Laflorencie}}]{colbois_interaction_2024}%
  \BibitemOpen
  \bibfield  {author} {\bibinfo {author} {\bibfnamefont {J.}~\bibnamefont {Colbois}}, \bibinfo {author} {\bibfnamefont {F.}~\bibnamefont {Alet}},\ and\ \bibinfo {author} {\bibfnamefont {N.}~\bibnamefont {Laflorencie}},\ }\bibfield  {title} {\bibinfo {title} {{Interaction-Driven Instabilities in the Random-Field XXZ Chain}},\ }\href {https://doi.org/10.1103/PhysRevLett.133.116502} {\bibfield  {journal} {\bibinfo  {journal} {Phys. Rev. Lett.}\ }\textbf {\bibinfo {volume} {133}},\ \bibinfo {pages} {116502} (\bibinfo {year} {2024}{\natexlab{a}})}\BibitemShut {NoStop}%
\bibitem [{\citenamefont {Suntajs}\ \emph {et~al.}(2020)\citenamefont {Suntajs}, \citenamefont {Bonca}, \citenamefont {Prosen},\ and\ \citenamefont {Vidmar}}]{suntajs_quantum_2020}%
  \BibitemOpen
  \bibfield  {author} {\bibinfo {author} {\bibfnamefont {J.}~\bibnamefont {Suntajs}}, \bibinfo {author} {\bibfnamefont {J.}~\bibnamefont {Bonca}}, \bibinfo {author} {\bibfnamefont {T.}~\bibnamefont {Prosen}},\ and\ \bibinfo {author} {\bibfnamefont {L.}~\bibnamefont {Vidmar}},\ }\bibfield  {title} {\bibinfo {title} {Quantum chaos challenges many-body localization},\ }\href {https://doi.org/10.1103/PhysRevE.102.062144} {\bibfield  {journal} {\bibinfo  {journal} {Phys. Rev. E}\ }\textbf {\bibinfo {volume} {102}},\ \bibinfo {pages} {062144} (\bibinfo {year} {2020})}\BibitemShut {NoStop}%
\bibitem [{\citenamefont {\ifmmode~\check{S}\else \v{S}\fi{}untajs}\ \emph {et~al.}(2020)\citenamefont {\ifmmode~\check{S}\else \v{S}\fi{}untajs}, \citenamefont {Bon\ifmmode~\check{c}\else \v{c}\fi{}a}, \citenamefont {Prosen},\ and\ \citenamefont {Vidmar}}]{suntajs_ergodicity_2020}%
  \BibitemOpen
  \bibfield  {author} {\bibinfo {author} {\bibfnamefont {J.}~\bibnamefont {\ifmmode~\check{S}\else \v{S}\fi{}untajs}}, \bibinfo {author} {\bibfnamefont {J.}~\bibnamefont {Bon\ifmmode~\check{c}\else \v{c}\fi{}a}}, \bibinfo {author} {\bibfnamefont {T.}~\bibnamefont {Prosen}},\ and\ \bibinfo {author} {\bibfnamefont {L.}~\bibnamefont {Vidmar}},\ }\bibfield  {title} {\bibinfo {title} {Ergodicity breaking transition in finite disordered spin chains},\ }\href {https://doi.org/10.1103/PhysRevB.102.064207} {\bibfield  {journal} {\bibinfo  {journal} {Phys. Rev. B}\ }\textbf {\bibinfo {volume} {102}},\ \bibinfo {pages} {064207} (\bibinfo {year} {2020})}\BibitemShut {NoStop}%
\bibitem [{\citenamefont {Sels}\ and\ \citenamefont {Polkovnikov}(2021)}]{sels_dynamical_2021}%
  \BibitemOpen
  \bibfield  {author} {\bibinfo {author} {\bibfnamefont {D.}~\bibnamefont {Sels}}\ and\ \bibinfo {author} {\bibfnamefont {A.}~\bibnamefont {Polkovnikov}},\ }\bibfield  {title} {\bibinfo {title} {Dynamical obstruction to localization in a disordered spin chain},\ }\href {https://doi.org/10.1103/PhysRevE.104.054105} {\bibfield  {journal} {\bibinfo  {journal} {Phys. Rev. E}\ }\textbf {\bibinfo {volume} {104}},\ \bibinfo {pages} {054105} (\bibinfo {year} {2021})}\BibitemShut {NoStop}%
\bibitem [{\citenamefont {Kiefer-Emmanouilidis}\ \emph {et~al.}(2022)\citenamefont {Kiefer-Emmanouilidis}, \citenamefont {Unanyan}, \citenamefont {Fleischhauer},\ and\ \citenamefont {Sirker}}]{sirker_particle_2022}%
  \BibitemOpen
  \bibfield  {author} {\bibinfo {author} {\bibfnamefont {M.}~\bibnamefont {Kiefer-Emmanouilidis}}, \bibinfo {author} {\bibfnamefont {R.}~\bibnamefont {Unanyan}}, \bibinfo {author} {\bibfnamefont {M.}~\bibnamefont {Fleischhauer}},\ and\ \bibinfo {author} {\bibfnamefont {J.}~\bibnamefont {Sirker}},\ }\bibfield  {title} {\bibinfo {title} {{Particle fluctuations and the failure of simple effective models for many-body localized phases}},\ }\href {https://doi.org/10.21468/SciPostPhys.12.1.034} {\bibfield  {journal} {\bibinfo  {journal} {SciPost Phys.}\ }\textbf {\bibinfo {volume} {12}},\ \bibinfo {pages} {034} (\bibinfo {year} {2022})}\BibitemShut {NoStop}%
\bibitem [{\citenamefont {Weisse}\ \emph {et~al.}(2025)\citenamefont {Weisse}, \citenamefont {Gerstner},\ and\ \citenamefont {Sirker}}]{weisse_operator_2025}%
  \BibitemOpen
  \bibfield  {author} {\bibinfo {author} {\bibfnamefont {A.}~\bibnamefont {Weisse}}, \bibinfo {author} {\bibfnamefont {R.}~\bibnamefont {Gerstner}},\ and\ \bibinfo {author} {\bibfnamefont {J.}~\bibnamefont {Sirker}},\ }\bibfield  {title} {\bibinfo {title} {Operator growth in disordered spin chains: Indications for the absence of many-body localization},\ }\href {https://doi.org/10.1103/wgss-nt8t} {\bibfield  {journal} {\bibinfo  {journal} {Phys. Rev. Res.}\ }\textbf {\bibinfo {volume} {7}},\ \bibinfo {pages} {033018} (\bibinfo {year} {2025})}\BibitemShut {NoStop}%
\bibitem [{\citenamefont {Sierant}\ \emph {et~al.}(2020{\natexlab{b}})\citenamefont {Sierant}, \citenamefont {Delande},\ and\ \citenamefont {Zakrzewski}}]{sierant_thouless_2020}%
  \BibitemOpen
  \bibfield  {author} {\bibinfo {author} {\bibfnamefont {P.}~\bibnamefont {Sierant}}, \bibinfo {author} {\bibfnamefont {D.}~\bibnamefont {Delande}},\ and\ \bibinfo {author} {\bibfnamefont {J.}~\bibnamefont {Zakrzewski}},\ }\bibfield  {title} {\bibinfo {title} {Thouless {Time} {Analysis} of {Anderson} and {Many}-{Body} {Localization} {Transitions}},\ }\href {https://doi.org/10.1103/PhysRevLett.124.186601} {\bibfield  {journal} {\bibinfo  {journal} {Phys. Rev. Lett.}\ }\textbf {\bibinfo {volume} {124}},\ \bibinfo {pages} {186601} (\bibinfo {year} {2020}{\natexlab{b}})}\BibitemShut {NoStop}%
\bibitem [{\citenamefont {Panda}\ \emph {et~al.}(2020)\citenamefont {Panda}, \citenamefont {Scardicchio}, \citenamefont {Schulz}, \citenamefont {Taylor},\ and\ \citenamefont {Žnidarič}}]{panda_can_2020}%
  \BibitemOpen
  \bibfield  {author} {\bibinfo {author} {\bibfnamefont {R.~K.}\ \bibnamefont {Panda}}, \bibinfo {author} {\bibfnamefont {A.}~\bibnamefont {Scardicchio}}, \bibinfo {author} {\bibfnamefont {M.}~\bibnamefont {Schulz}}, \bibinfo {author} {\bibfnamefont {S.~R.}\ \bibnamefont {Taylor}},\ and\ \bibinfo {author} {\bibfnamefont {M.}~\bibnamefont {Žnidarič}},\ }\bibfield  {title} {\bibinfo {title} {Can we study the many-body localisation transition {?}},\ }\href {https://doi.org/10.1209/0295-5075/128/67003} {\bibfield  {journal} {\bibinfo  {journal} {EPL}\ }\textbf {\bibinfo {volume} {128}},\ \bibinfo {pages} {67003} (\bibinfo {year} {2020})}\BibitemShut {NoStop}%
\bibitem [{\citenamefont {Abanin}\ \emph {et~al.}(2021)\citenamefont {Abanin}, \citenamefont {Bardarson}, \citenamefont {De~Tomasi}, \citenamefont {Gopalakrishnan}, \citenamefont {Khemani}, \citenamefont {Parameswaran}, \citenamefont {Pollmann}, \citenamefont {Potter}, \citenamefont {Serbyn},\ and\ \citenamefont {Vasseur}}]{abanin_distinguishing_2021}%
  \BibitemOpen
  \bibfield  {author} {\bibinfo {author} {\bibfnamefont {D.~A.}\ \bibnamefont {Abanin}}, \bibinfo {author} {\bibfnamefont {J.~H.}\ \bibnamefont {Bardarson}}, \bibinfo {author} {\bibfnamefont {G.}~\bibnamefont {De~Tomasi}}, \bibinfo {author} {\bibfnamefont {S.}~\bibnamefont {Gopalakrishnan}}, \bibinfo {author} {\bibfnamefont {V.}~\bibnamefont {Khemani}}, \bibinfo {author} {\bibfnamefont {S.~A.}\ \bibnamefont {Parameswaran}}, \bibinfo {author} {\bibfnamefont {F.}~\bibnamefont {Pollmann}}, \bibinfo {author} {\bibfnamefont {A.~C.}\ \bibnamefont {Potter}}, \bibinfo {author} {\bibfnamefont {M.}~\bibnamefont {Serbyn}},\ and\ \bibinfo {author} {\bibfnamefont {R.}~\bibnamefont {Vasseur}},\ }\bibfield  {title} {\bibinfo {title} {Distinguishing localization from chaos: {Challenges} in finite-size systems},\ }\href {https://doi.org/10.1016/j.aop.2021.168415} {\bibfield  {journal} {\bibinfo  {journal} {Annals of Physics}\ }\textbf {\bibinfo {volume} {427}},\ \bibinfo {pages} {168415} (\bibinfo {year} {2021})}\BibitemShut
  {NoStop}%
\bibitem [{\citenamefont {Sierant}\ and\ \citenamefont {Zakrzewski}(2022)}]{sierant_challenges_2022}%
  \BibitemOpen
  \bibfield  {author} {\bibinfo {author} {\bibfnamefont {P.}~\bibnamefont {Sierant}}\ and\ \bibinfo {author} {\bibfnamefont {J.}~\bibnamefont {Zakrzewski}},\ }\bibfield  {title} {\bibinfo {title} {Challenges to observation of many-body localization},\ }\href {https://doi.org/10.1103/PhysRevB.105.224203} {\bibfield  {journal} {\bibinfo  {journal} {Phys. Rev. B}\ }\textbf {\bibinfo {volume} {105}},\ \bibinfo {pages} {224203} (\bibinfo {year} {2022})}\BibitemShut {NoStop}%
\bibitem [{\citenamefont {Morningstar}\ \emph {et~al.}(2022)\citenamefont {Morningstar}, \citenamefont {Colmenarez}, \citenamefont {Khemani}, \citenamefont {Luitz},\ and\ \citenamefont {Huse}}]{morningstar_avalanches_2022}%
  \BibitemOpen
  \bibfield  {author} {\bibinfo {author} {\bibfnamefont {A.}~\bibnamefont {Morningstar}}, \bibinfo {author} {\bibfnamefont {L.}~\bibnamefont {Colmenarez}}, \bibinfo {author} {\bibfnamefont {V.}~\bibnamefont {Khemani}}, \bibinfo {author} {\bibfnamefont {D.~J.}\ \bibnamefont {Luitz}},\ and\ \bibinfo {author} {\bibfnamefont {D.~A.}\ \bibnamefont {Huse}},\ }\bibfield  {title} {\bibinfo {title} {Avalanches and many-body resonances in many-body localized systems},\ }\href {https://doi.org/10.1103/PhysRevB.105.174205} {\bibfield  {journal} {\bibinfo  {journal} {Phys. Rev. B}\ }\textbf {\bibinfo {volume} {105}},\ \bibinfo {pages} {174205} (\bibinfo {year} {2022})}\BibitemShut {NoStop}%
\bibitem [{\citenamefont {De~Roeck}\ and\ \citenamefont {Huveneers}(2017)}]{de_roeck_stability_2017}%
  \BibitemOpen
  \bibfield  {author} {\bibinfo {author} {\bibfnamefont {W.}~\bibnamefont {De~Roeck}}\ and\ \bibinfo {author} {\bibfnamefont {F.}~\bibnamefont {Huveneers}},\ }\bibfield  {title} {\bibinfo {title} {Stability and instability towards delocalization in many-body localization systems},\ }\href {https://doi.org/10.1103/PhysRevB.95.155129} {\bibfield  {journal} {\bibinfo  {journal} {Phys. Rev. B}\ }\textbf {\bibinfo {volume} {95}},\ \bibinfo {pages} {155129} (\bibinfo {year} {2017})}\BibitemShut {NoStop}%
\bibitem [{\citenamefont {Thiery}\ \emph {et~al.}(2018)\citenamefont {Thiery}, \citenamefont {Huveneers}, \citenamefont {M\"uller},\ and\ \citenamefont {De~Roeck}}]{thiery_many-body_2018}%
  \BibitemOpen
  \bibfield  {author} {\bibinfo {author} {\bibfnamefont {T.}~\bibnamefont {Thiery}}, \bibinfo {author} {\bibfnamefont {F.}~\bibnamefont {Huveneers}}, \bibinfo {author} {\bibfnamefont {M.}~\bibnamefont {M\"uller}},\ and\ \bibinfo {author} {\bibfnamefont {W.}~\bibnamefont {De~Roeck}},\ }\bibfield  {title} {\bibinfo {title} {Many-{Body} {Delocalization} as a {Quantum} {Avalanche}},\ }\href {https://doi.org/10.1103/PhysRevLett.121.140601} {\bibfield  {journal} {\bibinfo  {journal} {Phys. Rev. Lett.}\ }\textbf {\bibinfo {volume} {121}},\ \bibinfo {pages} {140601} (\bibinfo {year} {2018})}\BibitemShut {NoStop}%
\bibitem [{\citenamefont {Luitz}\ \emph {et~al.}(2017)\citenamefont {Luitz}, \citenamefont {Huveneers},\ and\ \citenamefont {De~Roeck}}]{luitz_how_2017}%
  \BibitemOpen
  \bibfield  {author} {\bibinfo {author} {\bibfnamefont {D.~J.}\ \bibnamefont {Luitz}}, \bibinfo {author} {\bibfnamefont {F.~m.~c.}\ \bibnamefont {Huveneers}},\ and\ \bibinfo {author} {\bibfnamefont {W.}~\bibnamefont {De~Roeck}},\ }\bibfield  {title} {\bibinfo {title} {{How a Small Quantum Bath Can Thermalize Long Localized Chains}},\ }\href {https://doi.org/10.1103/PhysRevLett.119.150602} {\bibfield  {journal} {\bibinfo  {journal} {Phys. Rev. Lett.}\ }\textbf {\bibinfo {volume} {119}},\ \bibinfo {pages} {150602} (\bibinfo {year} {2017})}\BibitemShut {NoStop}%
\bibitem [{\citenamefont {\ifmmode~\check{S}\else \v{S}\fi{}untajs}\ and\ \citenamefont {Vidmar}(2022)}]{suntajs_ergodicity_2022}%
  \BibitemOpen
  \bibfield  {author} {\bibinfo {author} {\bibfnamefont {J.}~\bibnamefont {\ifmmode~\check{S}\else \v{S}\fi{}untajs}}\ and\ \bibinfo {author} {\bibfnamefont {L.}~\bibnamefont {Vidmar}},\ }\bibfield  {title} {\bibinfo {title} {Ergodicity breaking transition in zero dimensions},\ }\href {https://doi.org/10.1103/PhysRevLett.129.060602} {\bibfield  {journal} {\bibinfo  {journal} {Phys. Rev. Lett.}\ }\textbf {\bibinfo {volume} {129}},\ \bibinfo {pages} {060602} (\bibinfo {year} {2022})}\BibitemShut {NoStop}%
\bibitem [{\citenamefont {Peacock}\ and\ \citenamefont {Sels}(2023)}]{peacock_many-body_2023}%
  \BibitemOpen
  \bibfield  {author} {\bibinfo {author} {\bibfnamefont {J.~C.}\ \bibnamefont {Peacock}}\ and\ \bibinfo {author} {\bibfnamefont {D.}~\bibnamefont {Sels}},\ }\bibfield  {title} {\bibinfo {title} {Many-body delocalization from embedded thermal inclusion},\ }\href {https://doi.org/10.1103/PhysRevB.108.L020201} {\bibfield  {journal} {\bibinfo  {journal} {Phys. Rev. B}\ }\textbf {\bibinfo {volume} {108}},\ \bibinfo {pages} {L020201} (\bibinfo {year} {2023})}\BibitemShut {NoStop}%
\bibitem [{\citenamefont {Ha}\ \emph {et~al.}(2023)\citenamefont {Ha}, \citenamefont {Morningstar},\ and\ \citenamefont {Huse}}]{ha_many-body_2023}%
  \BibitemOpen
  \bibfield  {author} {\bibinfo {author} {\bibfnamefont {H.}~\bibnamefont {Ha}}, \bibinfo {author} {\bibfnamefont {A.}~\bibnamefont {Morningstar}},\ and\ \bibinfo {author} {\bibfnamefont {D.~A.}\ \bibnamefont {Huse}},\ }\bibfield  {title} {\bibinfo {title} {Many-body resonances in the avalanche instability of many-body localization},\ }\href {https://doi.org/10.1103/PhysRevLett.130.250405} {\bibfield  {journal} {\bibinfo  {journal} {Phys. Rev. Lett.}\ }\textbf {\bibinfo {volume} {130}},\ \bibinfo {pages} {250405} (\bibinfo {year} {2023})}\BibitemShut {NoStop}%
\bibitem [{\citenamefont {Colmenarez}\ \emph {et~al.}(2024)\citenamefont {Colmenarez}, \citenamefont {Luitz},\ and\ \citenamefont {De~Roeck}}]{colmenarez_ergodic_2024}%
  \BibitemOpen
  \bibfield  {author} {\bibinfo {author} {\bibfnamefont {L.}~\bibnamefont {Colmenarez}}, \bibinfo {author} {\bibfnamefont {D.~J.}\ \bibnamefont {Luitz}},\ and\ \bibinfo {author} {\bibfnamefont {W.}~\bibnamefont {De~Roeck}},\ }\bibfield  {title} {\bibinfo {title} {Ergodic inclusions in many-body localized systems},\ }\href {https://doi.org/10.1103/PhysRevB.109.L081117} {\bibfield  {journal} {\bibinfo  {journal} {Phys. Rev. B}\ }\textbf {\bibinfo {volume} {109}},\ \bibinfo {pages} {L081117} (\bibinfo {year} {2024})}\BibitemShut {NoStop}%
\bibitem [{\citenamefont {Szo\l{}dra}\ \emph {et~al.}(2024)\citenamefont {Szo\l{}dra}, \citenamefont {Sierant}, \citenamefont {Lewenstein},\ and\ \citenamefont {Zakrzewski}}]{szoldra_catching_2024}%
  \BibitemOpen
  \bibfield  {author} {\bibinfo {author} {\bibfnamefont {T.}~\bibnamefont {Szo\l{}dra}}, \bibinfo {author} {\bibfnamefont {P.}~\bibnamefont {Sierant}}, \bibinfo {author} {\bibfnamefont {M.}~\bibnamefont {Lewenstein}},\ and\ \bibinfo {author} {\bibfnamefont {J.}~\bibnamefont {Zakrzewski}},\ }\bibfield  {title} {\bibinfo {title} {Catching thermal avalanches in the disordered {XXZ} model},\ }\href {https://doi.org/10.1103/PhysRevB.109.134202} {\bibfield  {journal} {\bibinfo  {journal} {Phys. Rev. B}\ }\textbf {\bibinfo {volume} {109}},\ \bibinfo {pages} {134202} (\bibinfo {year} {2024})}\BibitemShut {NoStop}%
\bibitem [{\citenamefont {Pawlik}\ \emph {et~al.}(2024)\citenamefont {Pawlik}, \citenamefont {Sierant}, \citenamefont {Vidmar},\ and\ \citenamefont {Zakrzewski}}]{pawlik_many-body_2024}%
  \BibitemOpen
  \bibfield  {author} {\bibinfo {author} {\bibfnamefont {K.}~\bibnamefont {Pawlik}}, \bibinfo {author} {\bibfnamefont {P.}~\bibnamefont {Sierant}}, \bibinfo {author} {\bibfnamefont {L.}~\bibnamefont {Vidmar}},\ and\ \bibinfo {author} {\bibfnamefont {J.}~\bibnamefont {Zakrzewski}},\ }\bibfield  {title} {\bibinfo {title} {Many-body mobility edge in quantum sun models},\ }\href {https://doi.org/10.1103/PhysRevB.109.L180201} {\bibfield  {journal} {\bibinfo  {journal} {Phys. Rev. B}\ }\textbf {\bibinfo {volume} {109}},\ \bibinfo {pages} {L180201} (\bibinfo {year} {2024})}\BibitemShut {NoStop}%
\bibitem [{\citenamefont {Pawlik}\ \emph {et~al.}()\citenamefont {Pawlik}, \citenamefont {Laflorencie},\ and\ \citenamefont {Zakrzewski}}]{pawlik2025unconventionalthermalizationlocalizedchain}%
  \BibitemOpen
  \bibfield  {author} {\bibinfo {author} {\bibfnamefont {K.}~\bibnamefont {Pawlik}}, \bibinfo {author} {\bibfnamefont {N.}~\bibnamefont {Laflorencie}},\ and\ \bibinfo {author} {\bibfnamefont {J.}~\bibnamefont {Zakrzewski}},\ }\href {https://arxiv.org/abs/2507.18286} {\bibinfo {title} {Unconventional thermalization of a localized chain interacting with an ergodic bath}},\ \Eprint {https://arxiv.org/abs/2507.18286} {arXiv:2507.18286} \BibitemShut {NoStop}%
\bibitem [{\citenamefont {Gopalakrishnan}\ \emph {et~al.}(2015)\citenamefont {Gopalakrishnan}, \citenamefont {M\"uller}, \citenamefont {Khemani}, \citenamefont {Knap}, \citenamefont {Demler},\ and\ \citenamefont {Huse}}]{gopalakrishnan_low_2015}%
  \BibitemOpen
  \bibfield  {author} {\bibinfo {author} {\bibfnamefont {S.}~\bibnamefont {Gopalakrishnan}}, \bibinfo {author} {\bibfnamefont {M.}~\bibnamefont {M\"uller}}, \bibinfo {author} {\bibfnamefont {V.}~\bibnamefont {Khemani}}, \bibinfo {author} {\bibfnamefont {M.}~\bibnamefont {Knap}}, \bibinfo {author} {\bibfnamefont {E.}~\bibnamefont {Demler}},\ and\ \bibinfo {author} {\bibfnamefont {D.~A.}\ \bibnamefont {Huse}},\ }\bibfield  {title} {\bibinfo {title} {Low-frequency conductivity in many-body localized systems},\ }\href {https://doi.org/10.1103/PhysRevB.92.104202} {\bibfield  {journal} {\bibinfo  {journal} {Phys. Rev. B}\ }\textbf {\bibinfo {volume} {92}},\ \bibinfo {pages} {104202} (\bibinfo {year} {2015})}\BibitemShut {NoStop}%
\bibitem [{\citenamefont {Kj\"all}(2018)}]{kjall_many-body_2018}%
  \BibitemOpen
  \bibfield  {author} {\bibinfo {author} {\bibfnamefont {J.~A.}\ \bibnamefont {Kj\"all}},\ }\bibfield  {title} {\bibinfo {title} {Many-body localization and level repulsion},\ }\href {https://doi.org/10.1103/PhysRevB.97.035163} {\bibfield  {journal} {\bibinfo  {journal} {Phys. Rev. B}\ }\textbf {\bibinfo {volume} {97}},\ \bibinfo {pages} {035163} (\bibinfo {year} {2018})}\BibitemShut {NoStop}%
\bibitem [{\citenamefont {Villalonga}\ and\ \citenamefont {Clark}(2020)}]{villalonga_eigenstates_2020}%
  \BibitemOpen
  \bibfield  {author} {\bibinfo {author} {\bibfnamefont {B.}~\bibnamefont {Villalonga}}\ and\ \bibinfo {author} {\bibfnamefont {B.~K.}\ \bibnamefont {Clark}},\ }\bibfield  {title} {\bibinfo {title} {Eigenstates hybridize on all length scales at the many-body localization transition},\ }\href {https://doi.org/10.48550/arXiv.2005.13558} {\bibfield  {journal} {\bibinfo  {journal} {arXiv:2005.13558}\ } (\bibinfo {year} {2020})}\BibitemShut {NoStop}%
\bibitem [{\citenamefont {Garratt}\ \emph {et~al.}(2021)\citenamefont {Garratt}, \citenamefont {Roy},\ and\ \citenamefont {Chalker}}]{garratt_local_2021}%
  \BibitemOpen
  \bibfield  {author} {\bibinfo {author} {\bibfnamefont {S.~J.}\ \bibnamefont {Garratt}}, \bibinfo {author} {\bibfnamefont {S.}~\bibnamefont {Roy}},\ and\ \bibinfo {author} {\bibfnamefont {J.~T.}\ \bibnamefont {Chalker}},\ }\bibfield  {title} {\bibinfo {title} {Local resonances and parametric level dynamics in the many-body localized phase},\ }\href {https://doi.org/10.1103/PhysRevB.104.184203} {\bibfield  {journal} {\bibinfo  {journal} {Phys. Rev. B}\ }\textbf {\bibinfo {volume} {104}},\ \bibinfo {pages} {184203} (\bibinfo {year} {2021})}\BibitemShut {NoStop}%
\bibitem [{\citenamefont {Garratt}\ and\ \citenamefont {Roy}(2022)}]{garratt_resonant_2022}%
  \BibitemOpen
  \bibfield  {author} {\bibinfo {author} {\bibfnamefont {S.~J.}\ \bibnamefont {Garratt}}\ and\ \bibinfo {author} {\bibfnamefont {S.}~\bibnamefont {Roy}},\ }\bibfield  {title} {\bibinfo {title} {Resonant energy scales and local observables in the many-body localized phase},\ }\href {https://doi.org/10.1103/PhysRevB.106.054309} {\bibfield  {journal} {\bibinfo  {journal} {Phys. Rev. B}\ }\textbf {\bibinfo {volume} {106}},\ \bibinfo {pages} {054309} (\bibinfo {year} {2022})}\BibitemShut {NoStop}%
\bibitem [{\citenamefont {Tikhonov}\ and\ \citenamefont {Mirlin}(2021{\natexlab{a}})}]{Tikhonov_Eigenstate_2021}%
  \BibitemOpen
  \bibfield  {author} {\bibinfo {author} {\bibfnamefont {K.~S.}\ \bibnamefont {Tikhonov}}\ and\ \bibinfo {author} {\bibfnamefont {A.~D.}\ \bibnamefont {Mirlin}},\ }\bibfield  {title} {\bibinfo {title} {Eigenstate correlations around the many-body localization transition},\ }\href {https://doi.org/10.1103/PhysRevB.103.064204} {\bibfield  {journal} {\bibinfo  {journal} {Phys. Rev. B}\ }\textbf {\bibinfo {volume} {103}},\ \bibinfo {pages} {064204} (\bibinfo {year} {2021}{\natexlab{a}})}\BibitemShut {NoStop}%
\bibitem [{\citenamefont {Tikhonov}\ and\ \citenamefont {Mirlin}(2021{\natexlab{b}})}]{Tikhonov_From_2021}%
  \BibitemOpen
  \bibfield  {author} {\bibinfo {author} {\bibfnamefont {K.}~\bibnamefont {Tikhonov}}\ and\ \bibinfo {author} {\bibfnamefont {A.}~\bibnamefont {Mirlin}},\ }\bibfield  {title} {\bibinfo {title} {From anderson localization on random regular graphs to many-body localization},\ }\href {https://doi.org/https://doi.org/10.1016/j.aop.2021.168525} {\bibfield  {journal} {\bibinfo  {journal} {Annals of Physics}\ }\textbf {\bibinfo {volume} {435}},\ \bibinfo {pages} {168525} (\bibinfo {year} {2021}{\natexlab{b}})},\ \bibinfo {note} {special Issue on Localisation 2020}\BibitemShut {NoStop}%
\bibitem [{\citenamefont {Crowley}\ and\ \citenamefont {Chandran}(2022)}]{crowley_constructive_2022}%
  \BibitemOpen
  \bibfield  {author} {\bibinfo {author} {\bibfnamefont {P.~J.~D.}\ \bibnamefont {Crowley}}\ and\ \bibinfo {author} {\bibfnamefont {A.}~\bibnamefont {Chandran}},\ }\bibfield  {title} {\bibinfo {title} {{A constructive theory of the numerically accessible many-body localized to thermal crossover}},\ }\href {https://doi.org/10.21468/SciPostPhys.12.6.201} {\bibfield  {journal} {\bibinfo  {journal} {SciPost Phys.}\ }\textbf {\bibinfo {volume} {12}},\ \bibinfo {pages} {201} (\bibinfo {year} {2022})}\BibitemShut {NoStop}%
\bibitem [{\citenamefont {Biroli}\ \emph {et~al.}(2024)\citenamefont {Biroli}, \citenamefont {Hartmann},\ and\ \citenamefont {Tarzia}}]{biroli_large-deviation_2024}%
  \BibitemOpen
  \bibfield  {author} {\bibinfo {author} {\bibfnamefont {G.}~\bibnamefont {Biroli}}, \bibinfo {author} {\bibfnamefont {A.~K.}\ \bibnamefont {Hartmann}},\ and\ \bibinfo {author} {\bibfnamefont {M.}~\bibnamefont {Tarzia}},\ }\bibfield  {title} {\bibinfo {title} {Large-deviation analysis of rare resonances for the many-body localization transition},\ }\href {https://doi.org/10.1103/PhysRevB.110.014205} {\bibfield  {journal} {\bibinfo  {journal} {Phys. Rev. B}\ }\textbf {\bibinfo {volume} {110}},\ \bibinfo {pages} {014205} (\bibinfo {year} {2024})}\BibitemShut {NoStop}%
\bibitem [{\citenamefont {Scoquart}\ \emph {et~al.}(2024)\citenamefont {Scoquart}, \citenamefont {Gornyi},\ and\ \citenamefont {Mirlin}}]{scoquart_role_2024}%
  \BibitemOpen
  \bibfield  {author} {\bibinfo {author} {\bibfnamefont {T.}~\bibnamefont {Scoquart}}, \bibinfo {author} {\bibfnamefont {I.~V.}\ \bibnamefont {Gornyi}},\ and\ \bibinfo {author} {\bibfnamefont {A.~D.}\ \bibnamefont {Mirlin}},\ }\bibfield  {title} {\bibinfo {title} {Role of fock-space correlations in many-body localization},\ }\href {https://doi.org/10.1103/PhysRevB.109.214203} {\bibfield  {journal} {\bibinfo  {journal} {Phys. Rev. B}\ }\textbf {\bibinfo {volume} {109}},\ \bibinfo {pages} {214203} (\bibinfo {year} {2024})}\BibitemShut {NoStop}%
\bibitem [{\citenamefont {Colbois}\ \emph {et~al.}(2024{\natexlab{b}})\citenamefont {Colbois}, \citenamefont {Alet},\ and\ \citenamefont {Laflorencie}}]{colbois_statistics_2024}%
  \BibitemOpen
  \bibfield  {author} {\bibinfo {author} {\bibfnamefont {J.}~\bibnamefont {Colbois}}, \bibinfo {author} {\bibfnamefont {F.}~\bibnamefont {Alet}},\ and\ \bibinfo {author} {\bibfnamefont {N.}~\bibnamefont {Laflorencie}},\ }\bibfield  {title} {\bibinfo {title} {Statistics of systemwide correlations in the random-field {XXZ} chain: Importance of rare events in the many-body localized phase},\ }\href {https://doi.org/10.1103/PhysRevB.110.214210} {\bibfield  {journal} {\bibinfo  {journal} {Phys. Rev. B}\ }\textbf {\bibinfo {volume} {110}},\ \bibinfo {pages} {214210} (\bibinfo {year} {2024}{\natexlab{b}})}\BibitemShut {NoStop}%
\bibitem [{\citenamefont {Laflorencie}\ \emph {et~al.}(2025)\citenamefont {Laflorencie}, \citenamefont {Colbois},\ and\ \citenamefont {Alet}}]{laflorencie_cat_2025}%
  \BibitemOpen
  \bibfield  {author} {\bibinfo {author} {\bibfnamefont {N.}~\bibnamefont {Laflorencie}}, \bibinfo {author} {\bibfnamefont {J.}~\bibnamefont {Colbois}},\ and\ \bibinfo {author} {\bibfnamefont {F.}~\bibnamefont {Alet}},\ }\bibfield  {title} {\bibinfo {title} {Cat states carrying long-range correlations in the many-body localized phase},\ }\href {https://doi.org/10.1103/gtfr-nblw} {\bibfield  {journal} {\bibinfo  {journal} {Phys. Rev. B}\ }\textbf {\bibinfo {volume} {112}},\ \bibinfo {pages} {224207} (\bibinfo {year} {2025})}\BibitemShut {NoStop}%
\bibitem [{\citenamefont {Harper}(1955)}]{harper_single_1955}%
  \BibitemOpen
  \bibfield  {author} {\bibinfo {author} {\bibfnamefont {P.~G.}\ \bibnamefont {Harper}},\ }\bibfield  {title} {\bibinfo {title} {Single {{Band Motion}} of {{Conduction Electrons}} in a {{Uniform Magnetic Field}}},\ }\href {https://doi.org/10.1088/0370-1298/68/10/304} {\bibfield  {journal} {\bibinfo  {journal} {Proc. Phys. Soc. A}\ }\textbf {\bibinfo {volume} {68}},\ \bibinfo {pages} {874} (\bibinfo {year} {1955})}\BibitemShut {NoStop}%
\bibitem [{\citenamefont {Aubry}\ and\ \citenamefont {Andr{\'e}}(1980)}]{AubryAndre}%
  \BibitemOpen
  \bibfield  {author} {\bibinfo {author} {\bibfnamefont {S.}~\bibnamefont {Aubry}}\ and\ \bibinfo {author} {\bibfnamefont {G.}~\bibnamefont {Andr{\'e}}},\ }\bibfield  {title} {\bibinfo {title} {Analyticity breaking and anderson localization in incommensurate lattices},\ }\href@noop {} {\bibfield  {journal} {\bibinfo  {journal} {Ann. Israel Phys. Soc}\ }\textbf {\bibinfo {volume} {3}},\ \bibinfo {pages} {18} (\bibinfo {year} {1980})}\BibitemShut {NoStop}%
\bibitem [{\citenamefont {Iyer}\ \emph {et~al.}(2013)\citenamefont {Iyer}, \citenamefont {Oganesyan}, \citenamefont {Refael},\ and\ \citenamefont {Huse}}]{Iyer_many-body_2013}%
  \BibitemOpen
  \bibfield  {author} {\bibinfo {author} {\bibfnamefont {S.}~\bibnamefont {Iyer}}, \bibinfo {author} {\bibfnamefont {V.}~\bibnamefont {Oganesyan}}, \bibinfo {author} {\bibfnamefont {G.}~\bibnamefont {Refael}},\ and\ \bibinfo {author} {\bibfnamefont {D.~A.}\ \bibnamefont {Huse}},\ }\bibfield  {title} {\bibinfo {title} {Many-body localization in a quasiperiodic system},\ }\href {https://doi.org/10.1103/PhysRevB.87.134202} {\bibfield  {journal} {\bibinfo  {journal} {Phys. Rev. B}\ }\textbf {\bibinfo {volume} {87}},\ \bibinfo {pages} {134202} (\bibinfo {year} {2013})}\BibitemShut {NoStop}%
\bibitem [{\citenamefont {Naldesi}\ \emph {et~al.}(2016)\citenamefont {Naldesi}, \citenamefont {Ercolessi},\ and\ \citenamefont {Roscilde}}]{Naldesi_2016}%
  \BibitemOpen
  \bibfield  {author} {\bibinfo {author} {\bibfnamefont {P.}~\bibnamefont {Naldesi}}, \bibinfo {author} {\bibfnamefont {E.}~\bibnamefont {Ercolessi}},\ and\ \bibinfo {author} {\bibfnamefont {T.}~\bibnamefont {Roscilde}},\ }\bibfield  {title} {\bibinfo {title} {{Detecting a many-body mobility edge with quantum quenches}},\ }\href {https://doi.org/10.21468/SciPostPhys.1.1.010} {\bibfield  {journal} {\bibinfo  {journal} {SciPost Phys.}\ }\textbf {\bibinfo {volume} {1}},\ \bibinfo {pages} {010} (\bibinfo {year} {2016})}\BibitemShut {NoStop}%
\bibitem [{\citenamefont {Setiawan}\ \emph {et~al.}(2017)\citenamefont {Setiawan}, \citenamefont {Deng},\ and\ \citenamefont {Pixley}}]{Setiawan_2017}%
  \BibitemOpen
  \bibfield  {author} {\bibinfo {author} {\bibfnamefont {F.}~\bibnamefont {Setiawan}}, \bibinfo {author} {\bibfnamefont {D.-L.}\ \bibnamefont {Deng}},\ and\ \bibinfo {author} {\bibfnamefont {J.~H.}\ \bibnamefont {Pixley}},\ }\bibfield  {title} {\bibinfo {title} {Transport properties across the many-body localization transition in quasiperiodic and random systems},\ }\href {https://doi.org/10.1103/PhysRevB.96.104205} {\bibfield  {journal} {\bibinfo  {journal} {Phys. Rev. B}\ }\textbf {\bibinfo {volume} {96}},\ \bibinfo {pages} {104205} (\bibinfo {year} {2017})}\BibitemShut {NoStop}%
\bibitem [{\citenamefont {Lee}\ \emph {et~al.}(2017)\citenamefont {Lee}, \citenamefont {Look}, \citenamefont {Lim},\ and\ \citenamefont {Sheng}}]{lee_many_2017}%
  \BibitemOpen
  \bibfield  {author} {\bibinfo {author} {\bibfnamefont {M.}~\bibnamefont {Lee}}, \bibinfo {author} {\bibfnamefont {T.~R.}\ \bibnamefont {Look}}, \bibinfo {author} {\bibfnamefont {S.~P.}\ \bibnamefont {Lim}},\ and\ \bibinfo {author} {\bibfnamefont {D.~N.}\ \bibnamefont {Sheng}},\ }\bibfield  {title} {\bibinfo {title} {Many-body localization in spin chain systems with quasiperiodic fields},\ }\href {https://doi.org/10.1103/PhysRevB.96.075146} {\bibfield  {journal} {\bibinfo  {journal} {Phys. Rev. B}\ }\textbf {\bibinfo {volume} {96}},\ \bibinfo {pages} {075146} (\bibinfo {year} {2017})}\BibitemShut {NoStop}%
\bibitem [{\citenamefont {Khemani}\ \emph {et~al.}(2017)\citenamefont {Khemani}, \citenamefont {Sheng},\ and\ \citenamefont {Huse}}]{Khemani_2017}%
  \BibitemOpen
  \bibfield  {author} {\bibinfo {author} {\bibfnamefont {V.}~\bibnamefont {Khemani}}, \bibinfo {author} {\bibfnamefont {D.~N.}\ \bibnamefont {Sheng}},\ and\ \bibinfo {author} {\bibfnamefont {D.~A.}\ \bibnamefont {Huse}},\ }\bibfield  {title} {\bibinfo {title} {Two universality classes for the many-body localization transition},\ }\href {https://doi.org/10.1103/PhysRevLett.119.075702} {\bibfield  {journal} {\bibinfo  {journal} {Phys. Rev. Lett.}\ }\textbf {\bibinfo {volume} {119}},\ \bibinfo {pages} {075702} (\bibinfo {year} {2017})}\BibitemShut {NoStop}%
\bibitem [{\citenamefont {Bera}\ \emph {et~al.}(2017)\citenamefont {Bera}, \citenamefont {De~Tomasi}, \citenamefont {Weiner},\ and\ \citenamefont {Evers}}]{Bera_2017}%
  \BibitemOpen
  \bibfield  {author} {\bibinfo {author} {\bibfnamefont {S.}~\bibnamefont {Bera}}, \bibinfo {author} {\bibfnamefont {G.}~\bibnamefont {De~Tomasi}}, \bibinfo {author} {\bibfnamefont {F.}~\bibnamefont {Weiner}},\ and\ \bibinfo {author} {\bibfnamefont {F.}~\bibnamefont {Evers}},\ }\bibfield  {title} {\bibinfo {title} {Density propagator for many-body localization: Finite-size effects, transient subdiffusion, and exponential decay},\ }\href {https://doi.org/10.1103/PhysRevLett.118.196801} {\bibfield  {journal} {\bibinfo  {journal} {Phys. Rev. Lett.}\ }\textbf {\bibinfo {volume} {118}},\ \bibinfo {pages} {196801} (\bibinfo {year} {2017})}\BibitemShut {NoStop}%
\bibitem [{\citenamefont {Nag}\ and\ \citenamefont {Garg}(2017)}]{nag_many_2017}%
  \BibitemOpen
  \bibfield  {author} {\bibinfo {author} {\bibfnamefont {S.}~\bibnamefont {Nag}}\ and\ \bibinfo {author} {\bibfnamefont {A.}~\bibnamefont {Garg}},\ }\bibfield  {title} {\bibinfo {title} {Many-body mobility edges in a one-dimensional system of interacting fermions},\ }\href {https://doi.org/10.1103/PhysRevB.96.060203} {\bibfield  {journal} {\bibinfo  {journal} {Phys. Rev. B}\ }\textbf {\bibinfo {volume} {96}},\ \bibinfo {pages} {060203} (\bibinfo {year} {2017})}\BibitemShut {NoStop}%
\bibitem [{\citenamefont {Bar~Lev}\ \emph {et~al.}(2017)\citenamefont {Bar~Lev}, \citenamefont {Kennes}, \citenamefont {Klöckner}, \citenamefont {Reichman},\ and\ \citenamefont {Karrasch}}]{bar_lev_transport_2017}%
  \BibitemOpen
  \bibfield  {author} {\bibinfo {author} {\bibfnamefont {Y.}~\bibnamefont {Bar~Lev}}, \bibinfo {author} {\bibfnamefont {D.~M.}\ \bibnamefont {Kennes}}, \bibinfo {author} {\bibfnamefont {C.}~\bibnamefont {Klöckner}}, \bibinfo {author} {\bibfnamefont {D.~R.}\ \bibnamefont {Reichman}},\ and\ \bibinfo {author} {\bibfnamefont {C.}~\bibnamefont {Karrasch}},\ }\bibfield  {title} {\bibinfo {title} {Transport in quasiperiodic interacting systems: From superdiffusion to subdiffusion},\ }\href {https://doi.org/10.1209/0295-5075/119/37003} {\bibfield  {journal} {\bibinfo  {journal} {Europhysics Letters}\ }\textbf {\bibinfo {volume} {119}},\ \bibinfo {pages} {37003} (\bibinfo {year} {2017})}\BibitemShut {NoStop}%
\bibitem [{\citenamefont {Žnidarič}\ and\ \citenamefont {Ljubotina}(2018)}]{_nidari__2018}%
  \BibitemOpen
  \bibfield  {author} {\bibinfo {author} {\bibfnamefont {M.}~\bibnamefont {Žnidarič}}\ and\ \bibinfo {author} {\bibfnamefont {M.}~\bibnamefont {Ljubotina}},\ }\bibfield  {title} {\bibinfo {title} {Interaction instability of localization in quasiperiodic systems},\ }\href {https://doi.org/10.1073/pnas.1800589115} {\bibfield  {journal} {\bibinfo  {journal} {Proceedings of the National Academy of Sciences}\ }\textbf {\bibinfo {volume} {115}},\ \bibinfo {pages} {4595} (\bibinfo {year} {2018})}\BibitemShut {NoStop}%
\bibitem [{\citenamefont {Weidinger}\ \emph {et~al.}(2018)\citenamefont {Weidinger}, \citenamefont {Gopalakrishnan},\ and\ \citenamefont {Knap}}]{Weidinger_2018}%
  \BibitemOpen
  \bibfield  {author} {\bibinfo {author} {\bibfnamefont {S.~A.}\ \bibnamefont {Weidinger}}, \bibinfo {author} {\bibfnamefont {S.}~\bibnamefont {Gopalakrishnan}},\ and\ \bibinfo {author} {\bibfnamefont {M.}~\bibnamefont {Knap}},\ }\bibfield  {title} {\bibinfo {title} {Self-consistent hartree-fock approach to many-body localization},\ }\href {https://doi.org/10.1103/PhysRevB.98.224205} {\bibfield  {journal} {\bibinfo  {journal} {Phys. Rev. B}\ }\textbf {\bibinfo {volume} {98}},\ \bibinfo {pages} {224205} (\bibinfo {year} {2018})}\BibitemShut {NoStop}%
\bibitem [{\citenamefont {Zhang}\ and\ \citenamefont {Yao}(2018)}]{Zhang_2018}%
  \BibitemOpen
  \bibfield  {author} {\bibinfo {author} {\bibfnamefont {S.-X.}\ \bibnamefont {Zhang}}\ and\ \bibinfo {author} {\bibfnamefont {H.}~\bibnamefont {Yao}},\ }\bibfield  {title} {\bibinfo {title} {Universal properties of many-body localization transitions in quasiperiodic systems},\ }\href {https://doi.org/10.1103/PhysRevLett.121.206601} {\bibfield  {journal} {\bibinfo  {journal} {Phys. Rev. Lett.}\ }\textbf {\bibinfo {volume} {121}},\ \bibinfo {pages} {206601} (\bibinfo {year} {2018})}\BibitemShut {NoStop}%
\bibitem [{\citenamefont {Doggen}\ and\ \citenamefont {Mirlin}(2019)}]{Doggen_2019}%
  \BibitemOpen
  \bibfield  {author} {\bibinfo {author} {\bibfnamefont {E.~V.~H.}\ \bibnamefont {Doggen}}\ and\ \bibinfo {author} {\bibfnamefont {A.~D.}\ \bibnamefont {Mirlin}},\ }\bibfield  {title} {\bibinfo {title} {Many-body delocalization dynamics in long aubry-andr\'e quasiperiodic chains},\ }\href {https://doi.org/10.1103/PhysRevB.100.104203} {\bibfield  {journal} {\bibinfo  {journal} {Phys. Rev. B}\ }\textbf {\bibinfo {volume} {100}},\ \bibinfo {pages} {104203} (\bibinfo {year} {2019})}\BibitemShut {NoStop}%
\bibitem [{\citenamefont {Xu}\ \emph {et~al.}(2019)\citenamefont {Xu}, \citenamefont {Li}, \citenamefont {Hsu}, \citenamefont {Swingle},\ and\ \citenamefont {Das~Sarma}}]{Xu_2019}%
  \BibitemOpen
  \bibfield  {author} {\bibinfo {author} {\bibfnamefont {S.}~\bibnamefont {Xu}}, \bibinfo {author} {\bibfnamefont {X.}~\bibnamefont {Li}}, \bibinfo {author} {\bibfnamefont {Y.-T.}\ \bibnamefont {Hsu}}, \bibinfo {author} {\bibfnamefont {B.}~\bibnamefont {Swingle}},\ and\ \bibinfo {author} {\bibfnamefont {S.}~\bibnamefont {Das~Sarma}},\ }\bibfield  {title} {\bibinfo {title} {Butterfly effect in interacting aubry-andre model: Thermalization, slow scrambling, and many-body localization},\ }\href {https://doi.org/10.1103/PhysRevResearch.1.032039} {\bibfield  {journal} {\bibinfo  {journal} {Phys. Rev. Res.}\ }\textbf {\bibinfo {volume} {1}},\ \bibinfo {pages} {032039} (\bibinfo {year} {2019})}\BibitemShut {NoStop}%
\bibitem [{\citenamefont {Weiner}\ \emph {et~al.}(2019)\citenamefont {Weiner}, \citenamefont {Evers},\ and\ \citenamefont {Bera}}]{Weiner_2019}%
  \BibitemOpen
  \bibfield  {author} {\bibinfo {author} {\bibfnamefont {F.}~\bibnamefont {Weiner}}, \bibinfo {author} {\bibfnamefont {F.}~\bibnamefont {Evers}},\ and\ \bibinfo {author} {\bibfnamefont {S.}~\bibnamefont {Bera}},\ }\bibfield  {title} {\bibinfo {title} {Slow dynamics and strong finite-size effects in many-body localization with random and quasiperiodic potentials},\ }\href {https://doi.org/10.1103/PhysRevB.100.104204} {\bibfield  {journal} {\bibinfo  {journal} {Phys. Rev. B}\ }\textbf {\bibinfo {volume} {100}},\ \bibinfo {pages} {104204} (\bibinfo {year} {2019})}\BibitemShut {NoStop}%
\bibitem [{\citenamefont {Sierant}\ and\ \citenamefont {Zakrzewski}(2019)}]{Sierant_2019}%
  \BibitemOpen
  \bibfield  {author} {\bibinfo {author} {\bibfnamefont {P.}~\bibnamefont {Sierant}}\ and\ \bibinfo {author} {\bibfnamefont {J.}~\bibnamefont {Zakrzewski}},\ }\bibfield  {title} {\bibinfo {title} {Level statistics across the many-body localization transition},\ }\href {https://doi.org/10.1103/PhysRevB.99.104205} {\bibfield  {journal} {\bibinfo  {journal} {Phys. Rev. B}\ }\textbf {\bibinfo {volume} {99}},\ \bibinfo {pages} {104205} (\bibinfo {year} {2019})}\BibitemShut {NoStop}%
\bibitem [{\citenamefont {Aramthottil}\ \emph {et~al.}(2021)\citenamefont {Aramthottil}, \citenamefont {Chanda}, \citenamefont {Sierant},\ and\ \citenamefont {Zakrzewski}}]{Aramthottil_2021}%
  \BibitemOpen
  \bibfield  {author} {\bibinfo {author} {\bibfnamefont {A.~S.}\ \bibnamefont {Aramthottil}}, \bibinfo {author} {\bibfnamefont {T.}~\bibnamefont {Chanda}}, \bibinfo {author} {\bibfnamefont {P.}~\bibnamefont {Sierant}},\ and\ \bibinfo {author} {\bibfnamefont {J.}~\bibnamefont {Zakrzewski}},\ }\bibfield  {title} {\bibinfo {title} {Finite-size scaling analysis of the many-body localization transition in quasiperiodic spin chains},\ }\href {https://doi.org/10.1103/PhysRevB.104.214201} {\bibfield  {journal} {\bibinfo  {journal} {Phys. Rev. B}\ }\textbf {\bibinfo {volume} {104}},\ \bibinfo {pages} {214201} (\bibinfo {year} {2021})}\BibitemShut {NoStop}%
\bibitem [{\citenamefont {Singh}\ \emph {et~al.}(2021)\citenamefont {Singh}, \citenamefont {Ware}, \citenamefont {Vasseur},\ and\ \citenamefont {Gopalakrishnan}}]{Singh_2021_qp}%
  \BibitemOpen
  \bibfield  {author} {\bibinfo {author} {\bibfnamefont {H.}~\bibnamefont {Singh}}, \bibinfo {author} {\bibfnamefont {B.}~\bibnamefont {Ware}}, \bibinfo {author} {\bibfnamefont {R.}~\bibnamefont {Vasseur}},\ and\ \bibinfo {author} {\bibfnamefont {S.}~\bibnamefont {Gopalakrishnan}},\ }\bibfield  {title} {\bibinfo {title} {Local integrals of motion and the quasiperiodic many-body localization transition},\ }\href {https://doi.org/10.1103/PhysRevB.103.L220201} {\bibfield  {journal} {\bibinfo  {journal} {Phys. Rev. B}\ }\textbf {\bibinfo {volume} {103}},\ \bibinfo {pages} {L220201} (\bibinfo {year} {2021})}\BibitemShut {NoStop}%
\bibitem [{\citenamefont {Agrawal}\ \emph {et~al.}(2022)\citenamefont {Agrawal}, \citenamefont {Vasseur},\ and\ \citenamefont {Gopalakrishnan}}]{Agrawal_2022}%
  \BibitemOpen
  \bibfield  {author} {\bibinfo {author} {\bibfnamefont {U.}~\bibnamefont {Agrawal}}, \bibinfo {author} {\bibfnamefont {R.}~\bibnamefont {Vasseur}},\ and\ \bibinfo {author} {\bibfnamefont {S.}~\bibnamefont {Gopalakrishnan}},\ }\bibfield  {title} {\bibinfo {title} {Quasiperiodic many-body localization transition in dimension $d>1$},\ }\href {https://doi.org/10.1103/PhysRevB.106.094206} {\bibfield  {journal} {\bibinfo  {journal} {Phys. Rev. B}\ }\textbf {\bibinfo {volume} {106}},\ \bibinfo {pages} {094206} (\bibinfo {year} {2022})}\BibitemShut {NoStop}%
\bibitem [{\citenamefont {\ifmmode~\check{S}\else \v{S}\fi{}trkalj}\ \emph {et~al.}(2022)\citenamefont {\ifmmode~\check{S}\else \v{S}\fi{}trkalj}, \citenamefont {Doggen},\ and\ \citenamefont {Castelnovo}}]{_trkalj_2022}%
  \BibitemOpen
  \bibfield  {author} {\bibinfo {author} {\bibfnamefont {A.}~\bibnamefont {\ifmmode~\check{S}\else \v{S}\fi{}trkalj}}, \bibinfo {author} {\bibfnamefont {E.~V.~H.}\ \bibnamefont {Doggen}},\ and\ \bibinfo {author} {\bibfnamefont {C.}~\bibnamefont {Castelnovo}},\ }\bibfield  {title} {\bibinfo {title} {Coexistence of localization and transport in many-body two-dimensional aubry-andr\'e models},\ }\href {https://doi.org/10.1103/PhysRevB.106.184209} {\bibfield  {journal} {\bibinfo  {journal} {Phys. Rev. B}\ }\textbf {\bibinfo {volume} {106}},\ \bibinfo {pages} {184209} (\bibinfo {year} {2022})}\BibitemShut {NoStop}%
\bibitem [{\citenamefont {Vu}\ \emph {et~al.}(2022)\citenamefont {Vu}, \citenamefont {Huang}, \citenamefont {Li},\ and\ \citenamefont {Das~Sarma}}]{Vu_2022}%
  \BibitemOpen
  \bibfield  {author} {\bibinfo {author} {\bibfnamefont {D.}~\bibnamefont {Vu}}, \bibinfo {author} {\bibfnamefont {K.}~\bibnamefont {Huang}}, \bibinfo {author} {\bibfnamefont {X.}~\bibnamefont {Li}},\ and\ \bibinfo {author} {\bibfnamefont {S.}~\bibnamefont {Das~Sarma}},\ }\bibfield  {title} {\bibinfo {title} {Fermionic many-body localization for random and quasiperiodic systems in the presence of short- and long-range interactions},\ }\href {https://doi.org/10.1103/PhysRevLett.128.146601} {\bibfield  {journal} {\bibinfo  {journal} {Phys. Rev. Lett.}\ }\textbf {\bibinfo {volume} {128}},\ \bibinfo {pages} {146601} (\bibinfo {year} {2022})}\BibitemShut {NoStop}%
\bibitem [{\citenamefont {Tu}\ \emph {et~al.}(2023{\natexlab{a}})\citenamefont {Tu}, \citenamefont {Vu},\ and\ \citenamefont {Das~Sarma}}]{Tu_2023}%
  \BibitemOpen
  \bibfield  {author} {\bibinfo {author} {\bibfnamefont {Y.-T.}\ \bibnamefont {Tu}}, \bibinfo {author} {\bibfnamefont {D.}~\bibnamefont {Vu}},\ and\ \bibinfo {author} {\bibfnamefont {S.}~\bibnamefont {Das~Sarma}},\ }\bibfield  {title} {\bibinfo {title} {Localization spectrum of a bath-coupled generalized aubry-andr\'e model in the presence of interactions},\ }\href {https://doi.org/10.1103/PhysRevB.108.064313} {\bibfield  {journal} {\bibinfo  {journal} {Phys. Rev. B}\ }\textbf {\bibinfo {volume} {108}},\ \bibinfo {pages} {064313} (\bibinfo {year} {2023}{\natexlab{a}})}\BibitemShut {NoStop}%
\bibitem [{\citenamefont {Thomson}\ and\ \citenamefont {Schirò}(2023)}]{Thomson_2023}%
  \BibitemOpen
  \bibfield  {author} {\bibinfo {author} {\bibfnamefont {S.~J.}\ \bibnamefont {Thomson}}\ and\ \bibinfo {author} {\bibfnamefont {M.}~\bibnamefont {Schirò}},\ }\bibfield  {title} {\bibinfo {title} {{Local integrals of motion in quasiperiodic many-body localized systems}},\ }\href {https://doi.org/10.21468/SciPostPhys.14.5.125} {\bibfield  {journal} {\bibinfo  {journal} {SciPost Phys.}\ }\textbf {\bibinfo {volume} {14}},\ \bibinfo {pages} {125} (\bibinfo {year} {2023})}\BibitemShut {NoStop}%
\bibitem [{\citenamefont {Prelov\ifmmode~\check{s}\else \v{s}\fi{}ek}\ \emph {et~al.}(2023)\citenamefont {Prelov\ifmmode~\check{s}\else \v{s}\fi{}ek}, \citenamefont {Herbrych},\ and\ \citenamefont {Mierzejewski}}]{Prelov_ek_2023}%
  \BibitemOpen
  \bibfield  {author} {\bibinfo {author} {\bibfnamefont {P.}~\bibnamefont {Prelov\ifmmode~\check{s}\else \v{s}\fi{}ek}}, \bibinfo {author} {\bibfnamefont {J.}~\bibnamefont {Herbrych}},\ and\ \bibinfo {author} {\bibfnamefont {M.}~\bibnamefont {Mierzejewski}},\ }\bibfield  {title} {\bibinfo {title} {Slow diffusion and thouless localization criterion in modulated spin chains},\ }\href {https://doi.org/10.1103/PhysRevB.108.035106} {\bibfield  {journal} {\bibinfo  {journal} {Phys. Rev. B}\ }\textbf {\bibinfo {volume} {108}},\ \bibinfo {pages} {035106} (\bibinfo {year} {2023})}\BibitemShut {NoStop}%
\bibitem [{\citenamefont {Falc\~ao}\ \emph {et~al.}(2024)\citenamefont {Falc\~ao}, \citenamefont {Aramthottil}, \citenamefont {Sierant},\ and\ \citenamefont {Zakrzewski}}]{Falc_o_2024}%
  \BibitemOpen
  \bibfield  {author} {\bibinfo {author} {\bibfnamefont {P.~R.~N.}\ \bibnamefont {Falc\~ao}}, \bibinfo {author} {\bibfnamefont {A.~S.}\ \bibnamefont {Aramthottil}}, \bibinfo {author} {\bibfnamefont {P.}~\bibnamefont {Sierant}},\ and\ \bibinfo {author} {\bibfnamefont {J.}~\bibnamefont {Zakrzewski}},\ }\bibfield  {title} {\bibinfo {title} {Many-body localization crossover is sharper in a quasiperiodic potential},\ }\href {https://doi.org/10.1103/PhysRevB.110.184209} {\bibfield  {journal} {\bibinfo  {journal} {Phys. Rev. B}\ }\textbf {\bibinfo {volume} {110}},\ \bibinfo {pages} {184209} (\bibinfo {year} {2024})}\BibitemShut {NoStop}%
\bibitem [{\citenamefont {Prasad}\ and\ \citenamefont {Garg}(2024)}]{Prasad_2024}%
  \BibitemOpen
  \bibfield  {author} {\bibinfo {author} {\bibfnamefont {Y.}~\bibnamefont {Prasad}}\ and\ \bibinfo {author} {\bibfnamefont {A.}~\bibnamefont {Garg}},\ }\bibfield  {title} {\bibinfo {title} {Single-particle excitations across the localization and many-body localization transition in quasiperiodic systems},\ }\href {https://doi.org/10.1103/PhysRevB.109.094204} {\bibfield  {journal} {\bibinfo  {journal} {Phys. Rev. B}\ }\textbf {\bibinfo {volume} {109}},\ \bibinfo {pages} {094204} (\bibinfo {year} {2024})}\BibitemShut {NoStop}%
\bibitem [{\citenamefont {Ghosh}\ \emph {et~al.}(2025)\citenamefont {Ghosh}, \citenamefont {Sutradhar}, \citenamefont {Mukerjee},\ and\ \citenamefont {Banerjee}}]{ghosh_scaling_2025}%
  \BibitemOpen
  \bibfield  {author} {\bibinfo {author} {\bibfnamefont {S.}~\bibnamefont {Ghosh}}, \bibinfo {author} {\bibfnamefont {J.}~\bibnamefont {Sutradhar}}, \bibinfo {author} {\bibfnamefont {S.}~\bibnamefont {Mukerjee}},\ and\ \bibinfo {author} {\bibfnamefont {S.}~\bibnamefont {Banerjee}},\ }\bibfield  {title} {\bibinfo {title} {Scaling of fock space propagator in quasiperiodic many-body localizing systems},\ }\href {https://doi.org/https://doi.org/10.1016/j.aop.2025.170001} {\bibfield  {journal} {\bibinfo  {journal} {Annals of Physics}\ }\textbf {\bibinfo {volume} {478}},\ \bibinfo {pages} {170001} (\bibinfo {year} {2025})}\BibitemShut {NoStop}%
\bibitem [{\citenamefont {Roati}\ \emph {et~al.}(2008)\citenamefont {Roati}, \citenamefont {D'Errico}, \citenamefont {Fallani}, \citenamefont {Fattori}, \citenamefont {Fort}, \citenamefont {Zaccanti}, \citenamefont {Modugno}, \citenamefont {Modugno},\ and\ \citenamefont {Inguscio}}]{roati_anderson_2008}%
  \BibitemOpen
  \bibfield  {author} {\bibinfo {author} {\bibfnamefont {G.}~\bibnamefont {Roati}}, \bibinfo {author} {\bibfnamefont {C.}~\bibnamefont {D'Errico}}, \bibinfo {author} {\bibfnamefont {L.}~\bibnamefont {Fallani}}, \bibinfo {author} {\bibfnamefont {M.}~\bibnamefont {Fattori}}, \bibinfo {author} {\bibfnamefont {C.}~\bibnamefont {Fort}}, \bibinfo {author} {\bibfnamefont {M.}~\bibnamefont {Zaccanti}}, \bibinfo {author} {\bibfnamefont {G.}~\bibnamefont {Modugno}}, \bibinfo {author} {\bibfnamefont {M.}~\bibnamefont {Modugno}},\ and\ \bibinfo {author} {\bibfnamefont {M.}~\bibnamefont {Inguscio}},\ }\bibfield  {title} {\bibinfo {title} {Anderson localization of a non-interacting {{Bose}}--{{Einstein}} condensate},\ }\href {https://doi.org/10.1038/nature07071} {\bibfield  {journal} {\bibinfo  {journal} {Nature}\ }\textbf {\bibinfo {volume} {453}},\ \bibinfo {pages} {895} (\bibinfo {year} {2008})}\BibitemShut {NoStop}%
\bibitem [{\citenamefont {Schreiber}\ \emph {et~al.}(2015)\citenamefont {Schreiber}, \citenamefont {Hodgman}, \citenamefont {Bordia}, \citenamefont {Lüschen}, \citenamefont {Fischer}, \citenamefont {Vosk}, \citenamefont {Altman}, \citenamefont {Schneider},\ and\ \citenamefont {Bloch}}]{Schreiber_2015}%
  \BibitemOpen
  \bibfield  {author} {\bibinfo {author} {\bibfnamefont {M.}~\bibnamefont {Schreiber}}, \bibinfo {author} {\bibfnamefont {S.~S.}\ \bibnamefont {Hodgman}}, \bibinfo {author} {\bibfnamefont {P.}~\bibnamefont {Bordia}}, \bibinfo {author} {\bibfnamefont {H.~P.}\ \bibnamefont {Lüschen}}, \bibinfo {author} {\bibfnamefont {M.~H.}\ \bibnamefont {Fischer}}, \bibinfo {author} {\bibfnamefont {R.}~\bibnamefont {Vosk}}, \bibinfo {author} {\bibfnamefont {E.}~\bibnamefont {Altman}}, \bibinfo {author} {\bibfnamefont {U.}~\bibnamefont {Schneider}},\ and\ \bibinfo {author} {\bibfnamefont {I.}~\bibnamefont {Bloch}},\ }\bibfield  {title} {\bibinfo {title} {Observation of many-body localization of interacting fermions in a quasirandom optical lattice},\ }\href {https://doi.org/10.1126/science.aaa7432} {\bibfield  {journal} {\bibinfo  {journal} {Science}\ }\textbf {\bibinfo {volume} {349}},\ \bibinfo {pages} {842} (\bibinfo {year} {2015})}\BibitemShut {NoStop}%
\bibitem [{\citenamefont {Bordia}\ \emph {et~al.}(2017)\citenamefont {Bordia}, \citenamefont {L\"uschen}, \citenamefont {Scherg}, \citenamefont {Gopalakrishnan}, \citenamefont {Knap}, \citenamefont {Schneider},\ and\ \citenamefont {Bloch}}]{bordia_probing_2017}%
  \BibitemOpen
  \bibfield  {author} {\bibinfo {author} {\bibfnamefont {P.}~\bibnamefont {Bordia}}, \bibinfo {author} {\bibfnamefont {H.}~\bibnamefont {L\"uschen}}, \bibinfo {author} {\bibfnamefont {S.}~\bibnamefont {Scherg}}, \bibinfo {author} {\bibfnamefont {S.}~\bibnamefont {Gopalakrishnan}}, \bibinfo {author} {\bibfnamefont {M.}~\bibnamefont {Knap}}, \bibinfo {author} {\bibfnamefont {U.}~\bibnamefont {Schneider}},\ and\ \bibinfo {author} {\bibfnamefont {I.}~\bibnamefont {Bloch}},\ }\bibfield  {title} {\bibinfo {title} {Probing slow relaxation and many-body localization in two-dimensional quasiperiodic systems},\ }\href {https://doi.org/10.1103/PhysRevX.7.041047} {\bibfield  {journal} {\bibinfo  {journal} {Phys. Rev. X}\ }\textbf {\bibinfo {volume} {7}},\ \bibinfo {pages} {041047} (\bibinfo {year} {2017})}\BibitemShut {NoStop}%
\bibitem [{\citenamefont {L\"uschen}\ \emph {et~al.}(2017)\citenamefont {L\"uschen}, \citenamefont {Bordia}, \citenamefont {Scherg}, \citenamefont {Alet}, \citenamefont {Altman}, \citenamefont {Schneider},\ and\ \citenamefont {Bloch}}]{L_schen_2017}%
  \BibitemOpen
  \bibfield  {author} {\bibinfo {author} {\bibfnamefont {H.~P.}\ \bibnamefont {L\"uschen}}, \bibinfo {author} {\bibfnamefont {P.}~\bibnamefont {Bordia}}, \bibinfo {author} {\bibfnamefont {S.}~\bibnamefont {Scherg}}, \bibinfo {author} {\bibfnamefont {F.}~\bibnamefont {Alet}}, \bibinfo {author} {\bibfnamefont {E.}~\bibnamefont {Altman}}, \bibinfo {author} {\bibfnamefont {U.}~\bibnamefont {Schneider}},\ and\ \bibinfo {author} {\bibfnamefont {I.}~\bibnamefont {Bloch}},\ }\bibfield  {title} {\bibinfo {title} {Observation of slow dynamics near the many-body localization transition in one-dimensional quasiperiodic systems},\ }\href {https://doi.org/10.1103/PhysRevLett.119.260401} {\bibfield  {journal} {\bibinfo  {journal} {Phys. Rev. Lett.}\ }\textbf {\bibinfo {volume} {119}},\ \bibinfo {pages} {260401} (\bibinfo {year} {2017})}\BibitemShut {NoStop}%
\bibitem [{\citenamefont {Lukin}\ \emph {et~al.}(2019)\citenamefont {Lukin}, \citenamefont {Rispoli}, \citenamefont {Schittko}, \citenamefont {Tai}, \citenamefont {Kaufman}, \citenamefont {Choi}, \citenamefont {Khemani}, \citenamefont {Léonard},\ and\ \citenamefont {Greiner}}]{Lukin_2019}%
  \BibitemOpen
  \bibfield  {author} {\bibinfo {author} {\bibfnamefont {A.}~\bibnamefont {Lukin}}, \bibinfo {author} {\bibfnamefont {M.}~\bibnamefont {Rispoli}}, \bibinfo {author} {\bibfnamefont {R.}~\bibnamefont {Schittko}}, \bibinfo {author} {\bibfnamefont {M.~E.}\ \bibnamefont {Tai}}, \bibinfo {author} {\bibfnamefont {A.~M.}\ \bibnamefont {Kaufman}}, \bibinfo {author} {\bibfnamefont {S.}~\bibnamefont {Choi}}, \bibinfo {author} {\bibfnamefont {V.}~\bibnamefont {Khemani}}, \bibinfo {author} {\bibfnamefont {J.}~\bibnamefont {Léonard}},\ and\ \bibinfo {author} {\bibfnamefont {M.}~\bibnamefont {Greiner}},\ }\bibfield  {title} {\bibinfo {title} {Probing entanglement in a many-body–localized system},\ }\href {https://doi.org/10.1126/science.aau0818} {\bibfield  {journal} {\bibinfo  {journal} {Science}\ }\textbf {\bibinfo {volume} {364}},\ \bibinfo {pages} {256} (\bibinfo {year} {2019})}\BibitemShut {NoStop}%
\bibitem [{\citenamefont {Rispoli}\ \emph {et~al.}(2019)\citenamefont {Rispoli}, \citenamefont {Lukin}, \citenamefont {Schittko}, \citenamefont {Kim}, \citenamefont {Tai}, \citenamefont {Léonard},\ and\ \citenamefont {Greiner}}]{Rispoli_2019}%
  \BibitemOpen
  \bibfield  {author} {\bibinfo {author} {\bibfnamefont {M.}~\bibnamefont {Rispoli}}, \bibinfo {author} {\bibfnamefont {A.}~\bibnamefont {Lukin}}, \bibinfo {author} {\bibfnamefont {R.}~\bibnamefont {Schittko}}, \bibinfo {author} {\bibfnamefont {S.}~\bibnamefont {Kim}}, \bibinfo {author} {\bibfnamefont {M.~E.}\ \bibnamefont {Tai}}, \bibinfo {author} {\bibfnamefont {J.}~\bibnamefont {Léonard}},\ and\ \bibinfo {author} {\bibfnamefont {M.}~\bibnamefont {Greiner}},\ }\bibfield  {title} {\bibinfo {title} {Quantum critical behaviour at the many-body localization transition},\ }\href {https://doi.org/10.1038/s41586-019-1527-2} {\bibfield  {journal} {\bibinfo  {journal} {Nature}\ }\textbf {\bibinfo {volume} {573}},\ \bibinfo {pages} {385} (\bibinfo {year} {2019})}\BibitemShut {NoStop}%
\bibitem [{\citenamefont {Léonard}\ \emph {et~al.}(2023)\citenamefont {Léonard}, \citenamefont {Kim}, \citenamefont {Rispoli}, \citenamefont {Lukin}, \citenamefont {Schittko}, \citenamefont {Kwan}, \citenamefont {Demler}, \citenamefont {Sels},\ and\ \citenamefont {Greiner}}]{L_onard_2023}%
  \BibitemOpen
  \bibfield  {author} {\bibinfo {author} {\bibfnamefont {J.}~\bibnamefont {Léonard}}, \bibinfo {author} {\bibfnamefont {S.}~\bibnamefont {Kim}}, \bibinfo {author} {\bibfnamefont {M.}~\bibnamefont {Rispoli}}, \bibinfo {author} {\bibfnamefont {A.}~\bibnamefont {Lukin}}, \bibinfo {author} {\bibfnamefont {R.}~\bibnamefont {Schittko}}, \bibinfo {author} {\bibfnamefont {J.}~\bibnamefont {Kwan}}, \bibinfo {author} {\bibfnamefont {E.}~\bibnamefont {Demler}}, \bibinfo {author} {\bibfnamefont {D.}~\bibnamefont {Sels}},\ and\ \bibinfo {author} {\bibfnamefont {M.}~\bibnamefont {Greiner}},\ }\bibfield  {title} {\bibinfo {title} {Probing the onset of quantum avalanches in a many-body localized system},\ }\href {https://doi.org/10.1038/s41567-022-01887-3} {\bibfield  {journal} {\bibinfo  {journal} {Nature Physics}\ }\textbf {\bibinfo {volume} {19}},\ \bibinfo {pages} {481} (\bibinfo {year} {2023})}\BibitemShut {NoStop}%
\bibitem [{\citenamefont {Hur}\ \emph {et~al.}()\citenamefont {Hur}, \citenamefont {Li}, \citenamefont {Lee}, \citenamefont {Kwon}, \citenamefont {Kim}, \citenamefont {Hwang}, \citenamefont {Kim}, \citenamefont {Yu}, \citenamefont {Chan}, \citenamefont {Wahl},\ and\ \citenamefont {yoon Choi}}]{hur2025stabilitymanybodylocalizationdimensions}%
  \BibitemOpen
  \bibfield  {author} {\bibinfo {author} {\bibfnamefont {J.}~\bibnamefont {Hur}}, \bibinfo {author} {\bibfnamefont {J.}~\bibnamefont {Li}}, \bibinfo {author} {\bibfnamefont {B.}~\bibnamefont {Lee}}, \bibinfo {author} {\bibfnamefont {K.}~\bibnamefont {Kwon}}, \bibinfo {author} {\bibfnamefont {M.}~\bibnamefont {Kim}}, \bibinfo {author} {\bibfnamefont {S.}~\bibnamefont {Hwang}}, \bibinfo {author} {\bibfnamefont {S.}~\bibnamefont {Kim}}, \bibinfo {author} {\bibfnamefont {Y.~S.}\ \bibnamefont {Yu}}, \bibinfo {author} {\bibfnamefont {A.}~\bibnamefont {Chan}}, \bibinfo {author} {\bibfnamefont {T.}~\bibnamefont {Wahl}},\ and\ \bibinfo {author} {\bibfnamefont {J.}~\bibnamefont {yoon Choi}},\ }\href {https://arxiv.org/abs/2508.20699} {\bibinfo {title} {Stability of many-body localization in two dimensions}},\ \Eprint {https://arxiv.org/abs/2508.20699} {arXiv:2508.20699} \BibitemShut {NoStop}%
\bibitem [{\citenamefont {{Li}}\ \emph {et~al.}(2023)\citenamefont {{Li}}, \citenamefont {{Wang}}, \citenamefont {{Shi}}, \citenamefont {{Huang}}, \citenamefont {{Song}}, \citenamefont {{Liang}}, \citenamefont {{Mei}}, \citenamefont {{Zhou}}, \citenamefont {{Zhang}}, \citenamefont {{Zhang}}, \citenamefont {{Chen}}, \citenamefont {{Zhao}}, \citenamefont {{Tian}}, \citenamefont {{Yang}}, \citenamefont {{Xiang}}, \citenamefont {{Xu}}, \citenamefont {{Zheng}},\ and\ \citenamefont {{Fan}}}]{li_observation_2023}%
  \BibitemOpen
  \bibfield  {author} {\bibinfo {author} {\bibfnamefont {H.}~\bibnamefont {{Li}}}, \bibinfo {author} {\bibfnamefont {Y.-Y.}\ \bibnamefont {{Wang}}}, \bibinfo {author} {\bibfnamefont {Y.-H.}\ \bibnamefont {{Shi}}}, \bibinfo {author} {\bibfnamefont {K.}~\bibnamefont {{Huang}}}, \bibinfo {author} {\bibfnamefont {X.}~\bibnamefont {{Song}}}, \bibinfo {author} {\bibfnamefont {G.-H.}\ \bibnamefont {{Liang}}}, \bibinfo {author} {\bibfnamefont {Z.-Y.}\ \bibnamefont {{Mei}}}, \bibinfo {author} {\bibfnamefont {B.}~\bibnamefont {{Zhou}}}, \bibinfo {author} {\bibfnamefont {H.}~\bibnamefont {{Zhang}}}, \bibinfo {author} {\bibfnamefont {J.-C.}\ \bibnamefont {{Zhang}}}, \bibinfo {author} {\bibfnamefont {S.}~\bibnamefont {{Chen}}}, \bibinfo {author} {\bibfnamefont {S.~P.}\ \bibnamefont {{Zhao}}}, \bibinfo {author} {\bibfnamefont {Y.}~\bibnamefont {{Tian}}}, \bibinfo {author} {\bibfnamefont {Z.-Y.}\ \bibnamefont {{Yang}}}, \bibinfo {author} {\bibfnamefont {Z.}~\bibnamefont {{Xiang}}}, \bibinfo {author} {\bibfnamefont
  {K.}~\bibnamefont {{Xu}}}, \bibinfo {author} {\bibfnamefont {D.}~\bibnamefont {{Zheng}}},\ and\ \bibinfo {author} {\bibfnamefont {H.}~\bibnamefont {{Fan}}},\ }\bibfield  {title} {\bibinfo {title} {{Observation of critical phase transition in a generalized Aubry-Andr{\'e}-Harper model with superconducting circuits}},\ }\href {https://doi.org/10.1038/s41534-023-00712-w} {\bibfield  {journal} {\bibinfo  {journal} {npj Quantum Information}\ }\textbf {\bibinfo {volume} {9}},\ \bibinfo {eid} {40} (\bibinfo {year} {2023})}\BibitemShut {NoStop}%
\bibitem [{\citenamefont {Tu}\ \emph {et~al.}(2023{\natexlab{b}})\citenamefont {Tu}, \citenamefont {Vu},\ and\ \citenamefont {Das~Sarma}}]{Tu_avalanche_2023}%
  \BibitemOpen
  \bibfield  {author} {\bibinfo {author} {\bibfnamefont {Y.-T.}\ \bibnamefont {Tu}}, \bibinfo {author} {\bibfnamefont {D.}~\bibnamefont {Vu}},\ and\ \bibinfo {author} {\bibfnamefont {S.}~\bibnamefont {Das~Sarma}},\ }\bibfield  {title} {\bibinfo {title} {Avalanche stability transition in interacting quasiperiodic systems},\ }\href {https://doi.org/10.1103/PhysRevB.107.014203} {\bibfield  {journal} {\bibinfo  {journal} {Phys. Rev. B}\ }\textbf {\bibinfo {volume} {107}},\ \bibinfo {pages} {014203} (\bibinfo {year} {2023}{\natexlab{b}})}\BibitemShut {NoStop}%
\bibitem [{\citenamefont {Agrawal}\ \emph {et~al.}(2020)\citenamefont {Agrawal}, \citenamefont {Gopalakrishnan},\ and\ \citenamefont {Vasseur}}]{Agrawal_2020}%
  \BibitemOpen
  \bibfield  {author} {\bibinfo {author} {\bibfnamefont {U.}~\bibnamefont {Agrawal}}, \bibinfo {author} {\bibfnamefont {S.}~\bibnamefont {Gopalakrishnan}},\ and\ \bibinfo {author} {\bibfnamefont {R.}~\bibnamefont {Vasseur}},\ }\bibfield  {title} {\bibinfo {title} {Universality and quantum criticality in quasiperiodic spin chains},\ }\href {https://doi.org/10.1038/s41467-020-15760-5} {\bibfield  {journal} {\bibinfo  {journal} {Nature Communications}\ }\textbf {\bibinfo {volume} {11}},\ \bibinfo {pages} {2225} (\bibinfo {year} {2020})}\BibitemShut {NoStop}%
\bibitem [{\citenamefont {Sun}\ \emph {et~al.}(2025)\citenamefont {Sun}, \citenamefont {Wang}, \citenamefont {Cui}, \citenamefont {Fan},\ and\ \citenamefont {Heyl}}]{sun_characterizing_2025}%
  \BibitemOpen
  \bibfield  {author} {\bibinfo {author} {\bibfnamefont {Z.-H.}\ \bibnamefont {Sun}}, \bibinfo {author} {\bibfnamefont {Y.-Y.}\ \bibnamefont {Wang}}, \bibinfo {author} {\bibfnamefont {J.}~\bibnamefont {Cui}}, \bibinfo {author} {\bibfnamefont {H.}~\bibnamefont {Fan}},\ and\ \bibinfo {author} {\bibfnamefont {M.}~\bibnamefont {Heyl}},\ }\bibfield  {title} {\bibinfo {title} {Characterizing dynamical criticality of many-body localization transitions from a {{Fock-space}} perspective},\ }\href {https://doi.org/10.1103/PhysRevB.111.094210} {\bibfield  {journal} {\bibinfo  {journal} {Phys. Rev. B}\ }\textbf {\bibinfo {volume} {111}},\ \bibinfo {pages} {094210} (\bibinfo {year} {2025})}\BibitemShut {NoStop}%
\bibitem [{\citenamefont {Giri}\ \emph {et~al.}(2025)\citenamefont {Giri}, \citenamefont {Siegl},\ and\ \citenamefont {Schliemann}}]{giri_from_2025}%
  \BibitemOpen
  \bibfield  {author} {\bibinfo {author} {\bibfnamefont {D.}~\bibnamefont {Giri}}, \bibinfo {author} {\bibfnamefont {J.}~\bibnamefont {Siegl}},\ and\ \bibinfo {author} {\bibfnamefont {J.}~\bibnamefont {Schliemann}},\ }\href {https://arxiv.org/abs/2508.00573} {\bibinfo {title} {From thermalization to multifractality: Spin-spin correlation in disordered $su(2)$-invariant 1d heisenberg spin chains}} (\bibinfo {year} {2025}),\ \Eprint {https://arxiv.org/abs/2508.00573} {arXiv:2508.00573} \BibitemShut {NoStop}%
\bibitem [{\citenamefont {Miranda}\ \emph {et~al.}(2025)\citenamefont {Miranda}, \citenamefont {Alet}, \citenamefont {Biroli}, \citenamefont {Cugliandolo}, \citenamefont {Laflorencie},\ and\ \citenamefont {Tarzia}}]{miranda_large-deviation_2025}%
  \BibitemOpen
  \bibfield  {author} {\bibinfo {author} {\bibfnamefont {G.~A.}\ \bibnamefont {Miranda}}, \bibinfo {author} {\bibfnamefont {F.}~\bibnamefont {Alet}}, \bibinfo {author} {\bibfnamefont {G.}~\bibnamefont {Biroli}}, \bibinfo {author} {\bibfnamefont {L.~F.}\ \bibnamefont {Cugliandolo}}, \bibinfo {author} {\bibfnamefont {N.}~\bibnamefont {Laflorencie}},\ and\ \bibinfo {author} {\bibfnamefont {M.}~\bibnamefont {Tarzia}},\ }\href {https://arxiv.org/abs/2510.18545} {\bibinfo {title} {Large deviations in the many-body localization transition: The case of the random-field xxz chain}} (\bibinfo {year} {2025}),\ \Eprint {https://arxiv.org/abs/2510.18545} {arXiv:2510.18545} \BibitemShut {NoStop}%
\bibitem [{Note1()}]{Note1}%
  \BibitemOpen
  \bibinfo {note} {Due to the Hamiltonian commuting with $S^z_{\protect \rm {total}}$, $x$ and $y$ components are equivalent by symmetry (with $\langle S_i^x\rangle =0$), while $\langle S_i^z\rangle \neq 0$ for longitudinal correlations.}\BibitemShut {Stop}%
\bibitem [{\citenamefont {Endres}\ \emph {et~al.}(2011)\citenamefont {Endres}, \citenamefont {Cheneau}, \citenamefont {Fukuhara}, \citenamefont {Weitenberg}, \citenamefont {Schauß}, \citenamefont {Gross}, \citenamefont {Mazza}, \citenamefont {Bañuls}, \citenamefont {Pollet}, \citenamefont {Bloch},\ and\ \citenamefont {Kuhr}}]{endres_observation_2011}%
  \BibitemOpen
  \bibfield  {author} {\bibinfo {author} {\bibfnamefont {M.}~\bibnamefont {Endres}}, \bibinfo {author} {\bibfnamefont {M.}~\bibnamefont {Cheneau}}, \bibinfo {author} {\bibfnamefont {T.}~\bibnamefont {Fukuhara}}, \bibinfo {author} {\bibfnamefont {C.}~\bibnamefont {Weitenberg}}, \bibinfo {author} {\bibfnamefont {P.}~\bibnamefont {Schauß}}, \bibinfo {author} {\bibfnamefont {C.}~\bibnamefont {Gross}}, \bibinfo {author} {\bibfnamefont {L.}~\bibnamefont {Mazza}}, \bibinfo {author} {\bibfnamefont {M.~C.}\ \bibnamefont {Bañuls}}, \bibinfo {author} {\bibfnamefont {L.}~\bibnamefont {Pollet}}, \bibinfo {author} {\bibfnamefont {I.}~\bibnamefont {Bloch}},\ and\ \bibinfo {author} {\bibfnamefont {S.}~\bibnamefont {Kuhr}},\ }\bibfield  {title} {\bibinfo {title} {Observation of correlated particle-hole pairs and string order in low-dimensional mott insulators},\ }\href {https://doi.org/10.1126/science.1209284} {\bibfield  {journal} {\bibinfo  {journal} {Science}\ }\textbf {\bibinfo {volume} {334}},\ \bibinfo {pages} {200}
  (\bibinfo {year} {2011})}\BibitemShut {NoStop}%
\bibitem [{\citenamefont {Bauer}\ and\ \citenamefont {Nayak}(2013)}]{bauer_area_2013}%
  \BibitemOpen
  \bibfield  {author} {\bibinfo {author} {\bibfnamefont {B.}~\bibnamefont {Bauer}}\ and\ \bibinfo {author} {\bibfnamefont {C.}~\bibnamefont {Nayak}},\ }\bibfield  {title} {\bibinfo {title} {Area laws in a many-body localized state and its implications for topological order},\ }\href {https://doi.org/10.1088/1742-5468/2013/09/P09005} {\bibfield  {journal} {\bibinfo  {journal} {Journal of Statistical Mechanics: Theory and Experiment}\ }\textbf {\bibinfo {volume} {2013}},\ \bibinfo {pages} {P09005} (\bibinfo {year} {2013})}\BibitemShut {NoStop}%
\bibitem [{SM()}]{SM}%
  \BibitemOpen
  \bibinfo {note} {See Supplemental Material at [URL] for additional analyses and technical details, which includes Ref.~[48].}\BibitemShut {Stop}%
\bibitem [{\citenamefont {Vidmar}\ and\ \citenamefont {Rigol}(2017)}]{vidmar_entanglement_2017}%
  \BibitemOpen
  \bibfield  {author} {\bibinfo {author} {\bibfnamefont {L.}~\bibnamefont {Vidmar}}\ and\ \bibinfo {author} {\bibfnamefont {M.}~\bibnamefont {Rigol}},\ }\bibfield  {title} {\bibinfo {title} {Entanglement entropy of eigenstates of quantum chaotic hamiltonians},\ }\href {https://doi.org/10.1103/PhysRevLett.119.220603} {\bibfield  {journal} {\bibinfo  {journal} {Phys. Rev. Lett.}\ }\textbf {\bibinfo {volume} {119}},\ \bibinfo {pages} {220603} (\bibinfo {year} {2017})}\BibitemShut {NoStop}%
\bibitem [{\citenamefont {Verstraten}\ \emph {et~al.}(2025)\citenamefont {Verstraten}, \citenamefont {Dai}, \citenamefont {Dixmerias}, \citenamefont {Peaudecerf}, \citenamefont {de~Jongh},\ and\ \citenamefont {Yefsah}}]{Tarik_correlations_2025}%
  \BibitemOpen
  \bibfield  {author} {\bibinfo {author} {\bibfnamefont {J.}~\bibnamefont {Verstraten}}, \bibinfo {author} {\bibfnamefont {K.}~\bibnamefont {Dai}}, \bibinfo {author} {\bibfnamefont {M.}~\bibnamefont {Dixmerias}}, \bibinfo {author} {\bibfnamefont {B.}~\bibnamefont {Peaudecerf}}, \bibinfo {author} {\bibfnamefont {T.}~\bibnamefont {de~Jongh}},\ and\ \bibinfo {author} {\bibfnamefont {T.}~\bibnamefont {Yefsah}},\ }\bibfield  {title} {\bibinfo {title} {In situ imaging of a single-atom wave packet in continuous space},\ }\href {https://doi.org/10.1103/PhysRevLett.134.083403} {\bibfield  {journal} {\bibinfo  {journal} {Phys. Rev. Lett.}\ }\textbf {\bibinfo {volume} {134}},\ \bibinfo {pages} {083403} (\bibinfo {year} {2025})}\BibitemShut {NoStop}%
\bibitem [{\citenamefont {Balay}\ \emph {et~al.}(2024)\citenamefont {Balay}, \citenamefont {Abhyankar}, \citenamefont {Adams}, \citenamefont {Benson}, \citenamefont {Brown}, \citenamefont {Brune}, \citenamefont {Buschelman}, \citenamefont {Constantinescu}, \citenamefont {Dalcin}, \citenamefont {Dener}, \citenamefont {Eijkhout}, \citenamefont {Faibussowitsch}, \citenamefont {Gropp}, \citenamefont {Hapla}, \citenamefont {Isaac}, \citenamefont {Jolivet}, \citenamefont {Karpeev}, \citenamefont {Kaushik}, \citenamefont {Knepley}, \citenamefont {Kong}, \citenamefont {Kruger}, \citenamefont {May}, \citenamefont {McInnes}, \citenamefont {Mills}, \citenamefont {Mitchell}, \citenamefont {Munson}, \citenamefont {Roman}, \citenamefont {Rupp}, \citenamefont {Sanan}, \citenamefont {Sarich}, \citenamefont {Smith}, \citenamefont {Zampini}, \citenamefont {Zhang}, \citenamefont {Zhang},\ and\ \citenamefont {Zhang}}]{petsc-user-ref}%
  \BibitemOpen
  \bibfield  {author} {\bibinfo {author} {\bibfnamefont {S.}~\bibnamefont {Balay}}, \bibinfo {author} {\bibfnamefont {S.}~\bibnamefont {Abhyankar}}, \bibinfo {author} {\bibfnamefont {M.~F.}\ \bibnamefont {Adams}}, \bibinfo {author} {\bibfnamefont {S.}~\bibnamefont {Benson}}, \bibinfo {author} {\bibfnamefont {J.}~\bibnamefont {Brown}}, \bibinfo {author} {\bibfnamefont {P.}~\bibnamefont {Brune}}, \bibinfo {author} {\bibfnamefont {K.}~\bibnamefont {Buschelman}}, \bibinfo {author} {\bibfnamefont {E.}~\bibnamefont {Constantinescu}}, \bibinfo {author} {\bibfnamefont {L.}~\bibnamefont {Dalcin}}, \bibinfo {author} {\bibfnamefont {A.}~\bibnamefont {Dener}}, \bibinfo {author} {\bibfnamefont {V.}~\bibnamefont {Eijkhout}}, \bibinfo {author} {\bibfnamefont {J.}~\bibnamefont {Faibussowitsch}}, \bibinfo {author} {\bibfnamefont {W.~D.}\ \bibnamefont {Gropp}}, \bibinfo {author} {\bibfnamefont {V.}~\bibnamefont {Hapla}}, \bibinfo {author} {\bibfnamefont {T.}~\bibnamefont {Isaac}}, \bibinfo {author} {\bibfnamefont
  {P.}~\bibnamefont {Jolivet}}, \bibinfo {author} {\bibfnamefont {D.}~\bibnamefont {Karpeev}}, \bibinfo {author} {\bibfnamefont {D.}~\bibnamefont {Kaushik}}, \bibinfo {author} {\bibfnamefont {M.~G.}\ \bibnamefont {Knepley}}, \bibinfo {author} {\bibfnamefont {F.}~\bibnamefont {Kong}}, \bibinfo {author} {\bibfnamefont {S.}~\bibnamefont {Kruger}}, \bibinfo {author} {\bibfnamefont {D.~A.}\ \bibnamefont {May}}, \bibinfo {author} {\bibfnamefont {L.~C.}\ \bibnamefont {McInnes}}, \bibinfo {author} {\bibfnamefont {R.~T.}\ \bibnamefont {Mills}}, \bibinfo {author} {\bibfnamefont {L.}~\bibnamefont {Mitchell}}, \bibinfo {author} {\bibfnamefont {T.}~\bibnamefont {Munson}}, \bibinfo {author} {\bibfnamefont {J.~E.}\ \bibnamefont {Roman}}, \bibinfo {author} {\bibfnamefont {K.}~\bibnamefont {Rupp}}, \bibinfo {author} {\bibfnamefont {P.}~\bibnamefont {Sanan}}, \bibinfo {author} {\bibfnamefont {J.}~\bibnamefont {Sarich}}, \bibinfo {author} {\bibfnamefont {B.~F.}\ \bibnamefont {Smith}}, \bibinfo {author} {\bibfnamefont
  {S.}~\bibnamefont {Zampini}}, \bibinfo {author} {\bibfnamefont {H.}~\bibnamefont {Zhang}}, \bibinfo {author} {\bibfnamefont {H.}~\bibnamefont {Zhang}},\ and\ \bibinfo {author} {\bibfnamefont {J.}~\bibnamefont {Zhang}},\ }\href {https://doi.org/10.2172/2205494} {\emph {\bibinfo {title} {{PETSc/TAO} Users Manual}}},\ \bibinfo {type} {Tech. Rep.}\ \bibinfo {number} {ANL-21/39 - Revision 3.21}\ (\bibinfo  {institution} {Argonne National Laboratory},\ \bibinfo {year} {2024})\BibitemShut {NoStop}%
\bibitem [{\citenamefont {Balay}\ \emph {et~al.}(1997)\citenamefont {Balay}, \citenamefont {Gropp}, \citenamefont {McInnes},\ and\ \citenamefont {Smith}}]{petsc-efficient}%
  \BibitemOpen
  \bibfield  {author} {\bibinfo {author} {\bibfnamefont {S.}~\bibnamefont {Balay}}, \bibinfo {author} {\bibfnamefont {W.~D.}\ \bibnamefont {Gropp}}, \bibinfo {author} {\bibfnamefont {L.~C.}\ \bibnamefont {McInnes}},\ and\ \bibinfo {author} {\bibfnamefont {B.~F.}\ \bibnamefont {Smith}},\ }\bibfield  {title} {\bibinfo {title} {Efficient management of parallelism in object oriented numerical software libraries},\ }in\ \href@noop {} {\emph {\bibinfo {booktitle} {Modern Software Tools in Scientific Computing}}},\ \bibinfo {editor} {edited by\ \bibinfo {editor} {\bibfnamefont {E.}~\bibnamefont {Arge}}, \bibinfo {editor} {\bibfnamefont {A.~M.}\ \bibnamefont {Bruaset}},\ and\ \bibinfo {editor} {\bibfnamefont {H.~P.}\ \bibnamefont {Langtangen}}}\ (\bibinfo  {publisher} {Birkh{\"{a}}user Press},\ \bibinfo {year} {1997})\ pp.\ \bibinfo {pages} {163--202}\BibitemShut {NoStop}%
\bibitem [{\citenamefont {Hernandez}\ \emph {et~al.}(2005{\natexlab{a}})\citenamefont {Hernandez}, \citenamefont {Roman},\ and\ \citenamefont {Vidal}}]{slepc-toms}%
  \BibitemOpen
  \bibfield  {author} {\bibinfo {author} {\bibfnamefont {V.}~\bibnamefont {Hernandez}}, \bibinfo {author} {\bibfnamefont {J.~E.}\ \bibnamefont {Roman}},\ and\ \bibinfo {author} {\bibfnamefont {V.}~\bibnamefont {Vidal}},\ }\bibfield  {title} {\bibinfo {title} {{SLEPc}: A scalable and flexible toolkit for the solution of eigenvalue problems},\ }\href {https://doi.org/https://doi.org/10.1145/1089014.1089019} {\bibfield  {journal} {\bibinfo  {journal} {{ACM} Trans. Math. Software}\ }\textbf {\bibinfo {volume} {31}},\ \bibinfo {pages} {351} (\bibinfo {year} {2005}{\natexlab{a}})}\BibitemShut {NoStop}%
\bibitem [{\citenamefont {Hernandez}\ \emph {et~al.}(2005{\natexlab{b}})\citenamefont {Hernandez}, \citenamefont {Roman},\ and\ \citenamefont {Vidal}}]{Hernandez:2005:SSF}%
  \BibitemOpen
  \bibfield  {author} {\bibinfo {author} {\bibfnamefont {V.}~\bibnamefont {Hernandez}}, \bibinfo {author} {\bibfnamefont {J.~E.}\ \bibnamefont {Roman}},\ and\ \bibinfo {author} {\bibfnamefont {V.}~\bibnamefont {Vidal}},\ }\bibfield  {title} {\bibinfo {title} {{SLEPc}: A scalable and flexible toolkit for the solution of eigenvalue problems},\ }\href@noop {} {\bibfield  {journal} {\bibinfo  {journal} {ACM Trans. Math. Software}\ }\textbf {\bibinfo {volume} {31}},\ \bibinfo {pages} {351} (\bibinfo {year} {2005}{\natexlab{b}})}\BibitemShut {NoStop}%
\bibitem [{\citenamefont {Roman}\ \emph {et~al.}(2023)\citenamefont {Roman}, \citenamefont {Campos}, \citenamefont {Dalcin}, \citenamefont {Romero},\ and\ \citenamefont {Tomas}}]{slepc-manual}%
  \BibitemOpen
  \bibfield  {author} {\bibinfo {author} {\bibfnamefont {J.~E.}\ \bibnamefont {Roman}}, \bibinfo {author} {\bibfnamefont {C.}~\bibnamefont {Campos}}, \bibinfo {author} {\bibfnamefont {L.}~\bibnamefont {Dalcin}}, \bibinfo {author} {\bibfnamefont {E.}~\bibnamefont {Romero}},\ and\ \bibinfo {author} {\bibfnamefont {A.}~\bibnamefont {Tomas}},\ }\href@noop {} {\emph {\bibinfo {title} {{SLEPc} Users Manual}}},\ \bibinfo {type} {Tech. Rep.}\ \bibinfo {number} {DSIC-II/24/02 - Revision 3.20}\ (\bibinfo  {institution} {D. Sistemes Inform\`atics i Computaci\'o, Universitat Polit\`ecnica de Val\`encia},\ \bibinfo {year} {2023})\BibitemShut {NoStop}%
\bibitem [{\citenamefont {Amestoy}\ \emph {et~al.}(2001)\citenamefont {Amestoy}, \citenamefont {Duff}, \citenamefont {Koster},\ and\ \citenamefont {L'Excellent}}]{MUMPS1}%
  \BibitemOpen
  \bibfield  {author} {\bibinfo {author} {\bibfnamefont {P.}~\bibnamefont {Amestoy}}, \bibinfo {author} {\bibfnamefont {I.~S.}\ \bibnamefont {Duff}}, \bibinfo {author} {\bibfnamefont {J.}~\bibnamefont {Koster}},\ and\ \bibinfo {author} {\bibfnamefont {J.-Y.}\ \bibnamefont {L'Excellent}},\ }\bibfield  {title} {\bibinfo {title} {A fully asynchronous multifrontal solver using distributed dynamic scheduling},\ }\href@noop {} {\bibfield  {journal} {\bibinfo  {journal} {SIAM Journal on Matrix Analysis and Applications}\ }\textbf {\bibinfo {volume} {23}},\ \bibinfo {pages} {15} (\bibinfo {year} {2001})}\BibitemShut {NoStop}%
\bibitem [{\citenamefont {Amestoy}\ \emph {et~al.}(2019)\citenamefont {Amestoy}, \citenamefont {Buttari}, \citenamefont {L'Excellent},\ and\ \citenamefont {Mary}}]{MUMPS2}%
  \BibitemOpen
  \bibfield  {author} {\bibinfo {author} {\bibfnamefont {P.}~\bibnamefont {Amestoy}}, \bibinfo {author} {\bibfnamefont {A.}~\bibnamefont {Buttari}}, \bibinfo {author} {\bibfnamefont {J.-Y.}\ \bibnamefont {L'Excellent}},\ and\ \bibinfo {author} {\bibfnamefont {T.}~\bibnamefont {Mary}},\ }\bibfield  {title} {\bibinfo {title} {{Performance and Scalability of the Block Low-Rank Multifrontal Factorization on Multicore Architectures}},\ }\href@noop {} {\bibfield  {journal} {\bibinfo  {journal} {ACM Transactions on Mathematical Software}\ }\textbf {\bibinfo {volume} {45}},\ \bibinfo {pages} {2:1} (\bibinfo {year} {2019})}\BibitemShut {NoStop}%
\bibitem [{\citenamefont {Ghysels}\ \emph {et~al.}(2016)\citenamefont {Ghysels}, \citenamefont {Li}, \citenamefont {Rouet}, \citenamefont {Williams},\ and\ \citenamefont {Napov}}]{Strumpack}%
  \BibitemOpen
  \bibfield  {author} {\bibinfo {author} {\bibfnamefont {P.}~\bibnamefont {Ghysels}}, \bibinfo {author} {\bibfnamefont {X.~S.}\ \bibnamefont {Li}}, \bibinfo {author} {\bibfnamefont {F.-H.}\ \bibnamefont {Rouet}}, \bibinfo {author} {\bibfnamefont {S.}~\bibnamefont {Williams}},\ and\ \bibinfo {author} {\bibfnamefont {A.}~\bibnamefont {Napov}},\ }\bibfield  {title} {\bibinfo {title} {An efficient multicore implementation of a novel hss-structured multifrontal solver using randomized sampling},\ }\href {https://doi.org/10.1137/15M1010117} {\bibfield  {journal} {\bibinfo  {journal} {SIAM Journal on Scientific Computing}\ }\textbf {\bibinfo {volume} {38}},\ \bibinfo {pages} {S358} (\bibinfo {year} {2016})}\BibitemShut {NoStop}%
\bibitem [{\citenamefont {Pietracaprina}\ \emph {et~al.}(2018)\citenamefont {Pietracaprina}, \citenamefont {Mac\'e}, \citenamefont {Luitz},\ and\ \citenamefont {Alet}}]{pietracaprina_shift-invert_2018}%
  \BibitemOpen
  \bibfield  {author} {\bibinfo {author} {\bibfnamefont {F.}~\bibnamefont {Pietracaprina}}, \bibinfo {author} {\bibfnamefont {N.}~\bibnamefont {Mac\'e}}, \bibinfo {author} {\bibfnamefont {D.~J.}\ \bibnamefont {Luitz}},\ and\ \bibinfo {author} {\bibfnamefont {F.}~\bibnamefont {Alet}},\ }\bibfield  {title} {\bibinfo {title} {Shift-invert diagonalization of large many-body localizing spin chains},\ }\href {https://scipost.org/SciPostPhys.5.5.045} {\bibfield  {journal} {\bibinfo  {journal} {SciPost Physics}\ }\textbf {\bibinfo {volume} {5}},\ \bibinfo {pages} {045} (\bibinfo {year} {2018})}\BibitemShut {NoStop}%
\bibitem [{\citenamefont {Rosenzweig}\ and\ \citenamefont {Porter}(1960)}]{rosenzweig_repulsion_1960}%
  \BibitemOpen
  \bibfield  {author} {\bibinfo {author} {\bibfnamefont {N.}~\bibnamefont {Rosenzweig}}\ and\ \bibinfo {author} {\bibfnamefont {C.~E.}\ \bibnamefont {Porter}},\ }\bibfield  {title} {\bibinfo {title} {{"Repulsion of Energy Levels" in Complex Atomic Spectra}},\ }\href {https://doi.org/10.1103/PhysRev.120.1698} {\bibfield  {journal} {\bibinfo  {journal} {Phys. Rev.}\ }\textbf {\bibinfo {volume} {120}},\ \bibinfo {pages} {1698} (\bibinfo {year} {1960})}\BibitemShut {NoStop}%
\bibitem [{\citenamefont {Oganesyan}\ and\ \citenamefont {Huse}(2007{\natexlab{b}})}]{oganesyan_localization_2007}%
  \BibitemOpen
  \bibfield  {author} {\bibinfo {author} {\bibfnamefont {V.}~\bibnamefont {Oganesyan}}\ and\ \bibinfo {author} {\bibfnamefont {D.~A.}\ \bibnamefont {Huse}},\ }\bibfield  {title} {\bibinfo {title} {Localization of interacting fermions at high temperature},\ }\href {https://doi.org/10.1103/PhysRevB.75.155111} {\bibfield  {journal} {\bibinfo  {journal} {Phys. Rev. B}\ }\textbf {\bibinfo {volume} {75}},\ \bibinfo {pages} {155111} (\bibinfo {year} {2007}{\natexlab{b}})}\BibitemShut {NoStop}%
\bibitem [{\citenamefont {Giraud}\ \emph {et~al.}(2022)\citenamefont {Giraud}, \citenamefont {Mace}, \citenamefont {Vernier},\ and\ \citenamefont {Alet}}]{giraud_probing_2022}%
  \BibitemOpen
  \bibfield  {author} {\bibinfo {author} {\bibfnamefont {O.}~\bibnamefont {Giraud}}, \bibinfo {author} {\bibfnamefont {N.}~\bibnamefont {Mace}}, \bibinfo {author} {\bibfnamefont {E.}~\bibnamefont {Vernier}},\ and\ \bibinfo {author} {\bibfnamefont {F.}~\bibnamefont {Alet}},\ }\bibfield  {title} {\bibinfo {title} {Probing {Symmetries} of {Quantum} {Many}-{Body} {Systems} through {Gap} {Ratio} {Statistics}},\ }\href {https://link.aps.org/doi/10.1103/PhysRevX.12.011006} {\bibfield  {journal} {\bibinfo  {journal} {Phys. Rev. X}\ }\textbf {\bibinfo {volume} {12}},\ \bibinfo {pages} {011006} (\bibinfo {year} {2022})}\BibitemShut {NoStop}%
\bibitem [{\citenamefont {Kullback}\ and\ \citenamefont {Leibler}(1951)}]{kullback_information_1951}%
  \BibitemOpen
  \bibfield  {author} {\bibinfo {author} {\bibfnamefont {S.}~\bibnamefont {Kullback}}\ and\ \bibinfo {author} {\bibfnamefont {R.~A.}\ \bibnamefont {Leibler}},\ }\bibfield  {title} {\bibinfo {title} {{On Information and Sufficiency}},\ }\href {https://doi.org/10.1214/aoms/1177729694} {\bibfield  {journal} {\bibinfo  {journal} {Ann. Math. Statist.}\ }\textbf {\bibinfo {volume} {22}},\ \bibinfo {pages} {79} (\bibinfo {year} {1951})}\BibitemShut {NoStop}%
\bibitem [{\citenamefont {Luca}\ and\ \citenamefont {Scardicchio}(2013)}]{luca_ergodicity_2013}%
  \BibitemOpen
  \bibfield  {author} {\bibinfo {author} {\bibfnamefont {A.~D.}\ \bibnamefont {Luca}}\ and\ \bibinfo {author} {\bibfnamefont {A.}~\bibnamefont {Scardicchio}},\ }\bibfield  {title} {\bibinfo {title} {Ergodicity breaking in a model showing many-body localization},\ }\href {https://doi.org/10.1209/0295-5075/101/37003} {\bibfield  {journal} {\bibinfo  {journal} {EPL}\ }\textbf {\bibinfo {volume} {101}},\ \bibinfo {pages} {37003} (\bibinfo {year} {2013})}\BibitemShut {NoStop}%
\bibitem [{\citenamefont {Mac\'e}\ \emph {et~al.}(2019)\citenamefont {Mac\'e}, \citenamefont {Alet},\ and\ \citenamefont {Laflorencie}}]{mace_multifractal_2019}%
  \BibitemOpen
  \bibfield  {author} {\bibinfo {author} {\bibfnamefont {N.}~\bibnamefont {Mac\'e}}, \bibinfo {author} {\bibfnamefont {F.}~\bibnamefont {Alet}},\ and\ \bibinfo {author} {\bibfnamefont {N.}~\bibnamefont {Laflorencie}},\ }\bibfield  {title} {\bibinfo {title} {Multifractal {Scalings} {Across} the {Many}-{Body} {Localization} {Transition}},\ }\href {https://doi.org/10.1103/PhysRevLett.123.180601} {\bibfield  {journal} {\bibinfo  {journal} {Phys. Rev. Lett.}\ }\textbf {\bibinfo {volume} {123}},\ \bibinfo {pages} {180601} (\bibinfo {year} {2019})}\BibitemShut {NoStop}%
\bibitem [{\citenamefont {Laflorencie}\ \emph {et~al.}(2020)\citenamefont {Laflorencie}, \citenamefont {Lemari\'e},\ and\ \citenamefont {Mac\'e}}]{laflorencie_chain_2020}%
  \BibitemOpen
  \bibfield  {author} {\bibinfo {author} {\bibfnamefont {N.}~\bibnamefont {Laflorencie}}, \bibinfo {author} {\bibfnamefont {G.}~\bibnamefont {Lemari\'e}},\ and\ \bibinfo {author} {\bibfnamefont {N.}~\bibnamefont {Mac\'e}},\ }\bibfield  {title} {\bibinfo {title} {Chain breaking and {Kosterlitz}-{Thouless} scaling at the many-body localization transition in the random-field {Heisenberg} spin chain},\ }\href {https://doi.org/10.1103/PhysRevResearch.2.042033} {\bibfield  {journal} {\bibinfo  {journal} {Phys. Rev. Research}\ }\textbf {\bibinfo {volume} {2}},\ \bibinfo {pages} {042033(R)} (\bibinfo {year} {2020})}\BibitemShut {NoStop}%
\bibitem [{\citenamefont {Colbois}\ and\ \citenamefont {Laflorencie}(2023)}]{colbois_breaking_2023}%
  \BibitemOpen
  \bibfield  {author} {\bibinfo {author} {\bibfnamefont {J.}~\bibnamefont {Colbois}}\ and\ \bibinfo {author} {\bibfnamefont {N.}~\bibnamefont {Laflorencie}},\ }\bibfield  {title} {\bibinfo {title} {Breaking the chains: {Extreme} value statistics and localization in random spin chains},\ }\href {https://doi.org/10.1103/PhysRevB.108.144206} {\bibfield  {journal} {\bibinfo  {journal} {Phys. Rev. B}\ }\textbf {\bibinfo {volume} {108}},\ \bibinfo {pages} {144206} (\bibinfo {year} {2023})}\BibitemShut {NoStop}%
\bibitem [{\citenamefont {Sai~Aramthottil}\ \emph {et~al.}(2025)\citenamefont {Sai~Aramthottil}, \citenamefont {Emami~Kopaei}, \citenamefont {Sierant}, \citenamefont {Vidmar},\ and\ \citenamefont {Zakrzewski}}]{aramthottil_false_2025}%
  \BibitemOpen
  \bibfield  {author} {\bibinfo {author} {\bibfnamefont {A.}~\bibnamefont {Sai~Aramthottil}}, \bibinfo {author} {\bibfnamefont {A.}~\bibnamefont {Emami~Kopaei}}, \bibinfo {author} {\bibfnamefont {P.}~\bibnamefont {Sierant}}, \bibinfo {author} {\bibfnamefont {L.}~\bibnamefont {Vidmar}},\ and\ \bibinfo {author} {\bibfnamefont {J.}~\bibnamefont {Zakrzewski}},\ }\bibfield  {title} {\bibinfo {title} {False signatures of non-ergodic behavior in disordered quantum many-body systems},\ }\href {https://doi.org/10.1088/1367-2630/ae20b3} {\bibfield  {journal} {\bibinfo  {journal} {New Journal of Physics}\ }\textbf {\bibinfo {volume} {27}},\ \bibinfo {pages} {125001} (\bibinfo {year} {2025})}\BibitemShut {NoStop}%
\end{thebibliography}%

\onecolumngrid

\vspace{1cm}

\begin{center}
    \bfseries\large End Matter
\end{center}

\twocolumngrid

\section{Simulation details}
All the observables for this study are evaluated in many-body eigenstates near the center of the spectrum, i.e, at density $\epsilon = 0.5$ with the energy density defined as
$\epsilon = (E - E_{\min})/(E_{\max} - E_{\min})$
where $E$ is the eigenenergy, and $E_{\min}$ ($E_{\max}$) denotes the minimum (maximum) many-body energy within the symmetry sector. We restrict to the zero magnetization sector, $S^z_\mathrm{total} = 0$ with Hilbert space dimension $\mathcal{N}=L!/[(L/2)!]^2$. We employ shift-invert exact diagonalization to access highly excited states in systems of size up to $L = 22$ spins~\cite{pietracaprina_shift-invert_2018,luitz_many-body_2015}.
\label{sec:simulation_details}
Tab.~\ref{table:Samples} summarizes the parameters used in the shift-invert computation for each system size $L$.  

\begin{table}[ht]
    \centering
    \begin{tabular}{c c c|c c c}
    \toprule
    \textbf{$L$} & \textbf{$N_{\mathrm{eig}}$} & \textbf{$N_{\mathrm{samp}}$} & \textbf{$L$} & \textbf{$N_{\mathrm{eig}}$} & \textbf{$N_{\mathrm{samp}}$} \\
    \midrule
    10 & $\min(10, N_{0.005})$ & 8000 & 18 & $\min(60, N_{0.005})$ & 1500 \\
    12 & $\min(30, N_{0.005})$ & 4000 & 20 & $\min(60, N_{0.005})$ & 1000 \\
    14 & $\min(60, N_{0.005})$ & 2500 & 22 & $\min(60, N_{0.005})$ & 1000 \\
    16 & $\min(60, N_{0.005})$ & 2500 & ~ & ~ & ~ \\
    \bottomrule
    \end{tabular}
    \caption{
    Number of eigenstates ($N_{\mathrm{eig}}$) and QP samples ($N_{\mathrm{samp}}$) used for each system size. 
    For each $L$, $N_{\mathrm{eig}}$ eigenstates are taken from the middle of the spectrum within an energy window of $0.005$ ($N_{0.005}$).
    }
    \label{table:Samples}
\end{table}

\section{Observables}
\label{EM:probes}
This section details the key observables used in the main text to characterize the quasiperiodic Heisenberg chain.

\textit{\textbf{Gap Ratio (GR).}}
The gap ratio (GR) is a key diagnostic for distinguishing thermal from localized phases by analyzing spectral statistics~\cite{rosenzweig_repulsion_1960,jacquod_emergence_1997,oganesyan_localization_2007,Pal_2010,alet_many-body_2018,giraud_probing_2022}. It is defined as~\cite{oganesyan_localization_2007}
\begin{equation}
r = \frac{\min(\delta_n, \delta_{n+1})}{\max(\delta_n, \delta_{n+1})},
\end{equation}
where $\delta_n = E_{n+1} - E_n$ are consecutive energy level spacings. Infinite-temperature states as modeled by random states (in the Gaussian Orthogonal Ensemble, GOE) exhibit level repulsion, while  uncorrelated levels in MBL phases (or integrable systems) show Poisson statistics, with distributions
\begin{align}
P_{\mathrm{GOE}}(r) &= \frac{27}{4} \frac{r + r^2}{(1 + r + r^2)^{5/2}}, \quad r \in [0, 1], \\
P_{\mathrm{Poisson}}(r) &= \frac{2}{(1 + r)^2}, \quad r \in [0, 1].
\end{align}
In the scaling analysis to extract $h_{\rm c}^{\rm standard}$, we use the crossing of the Kullback–Leibler (KL)~\cite{kullback_information_1951} divergence between the numerically evaluated $P(r)$ and $P_{\mathrm{Poisson}}(r)$ as a robust quantitative measure showing clearer contrast between the Poisson and GOE distributions~\cite{giraud_probing_2022,colbois_interaction_2024}
\begin{equation}
\text{KL}_{P(r) | P_{\text{Poisson}}(r)} = \int_0^1 P(r) \ln \left( \frac{P(r)}{P_{\text{Poisson}}(r)} \right) dr,
\end{equation}
with limits $\text{KL}_{P(r) | P_{\text{Poisson}}(r)} = 0$ for the localized case and $\text{KL}_{P_\text{GOE}(r) | P_{\text{Poisson}}(r)}\approx 0.1895$ for the ergodic case.\\

\textit{\textbf{Entanglement Entropy (EE).}} The entanglement entropy $S_A$ for a subsystem $A$ is defined as $S_A = -\text{Tr} (\rho_A \ln \rho_A)$, where $\rho_A = \text{Tr}_B (|\psi\rangle\langle\psi|)$ is the reduced density matrix. Eigenstates of thermal or ergodic phases exhibit volume-law scaling of $S_A$ following random matrix theory~\cite{vidmar_entanglement_2017,Aramthottil_2021} $S_{\rm RMT} = \frac{1}{2}(L \ln (2) +  \ln{1/2}-1)$ (in the $S^z_{\rm total} = 0$ sector), while MBL phases show area-law scaling~\cite{bauer_area_2013,luitz_many-body_2015}, indicating low entanglement due to localized eigenstates. We compute the half-chain EE (i.e., subsystem $A$ is made of $L/2$ consecutive sites) and average over all cuts, all obtained eigenstates, and QP realizations (Tab.~\ref{table:Samples}).\\
\begin{figure*}[t!]
    \centering
    \includegraphics[width=0.85\textwidth]{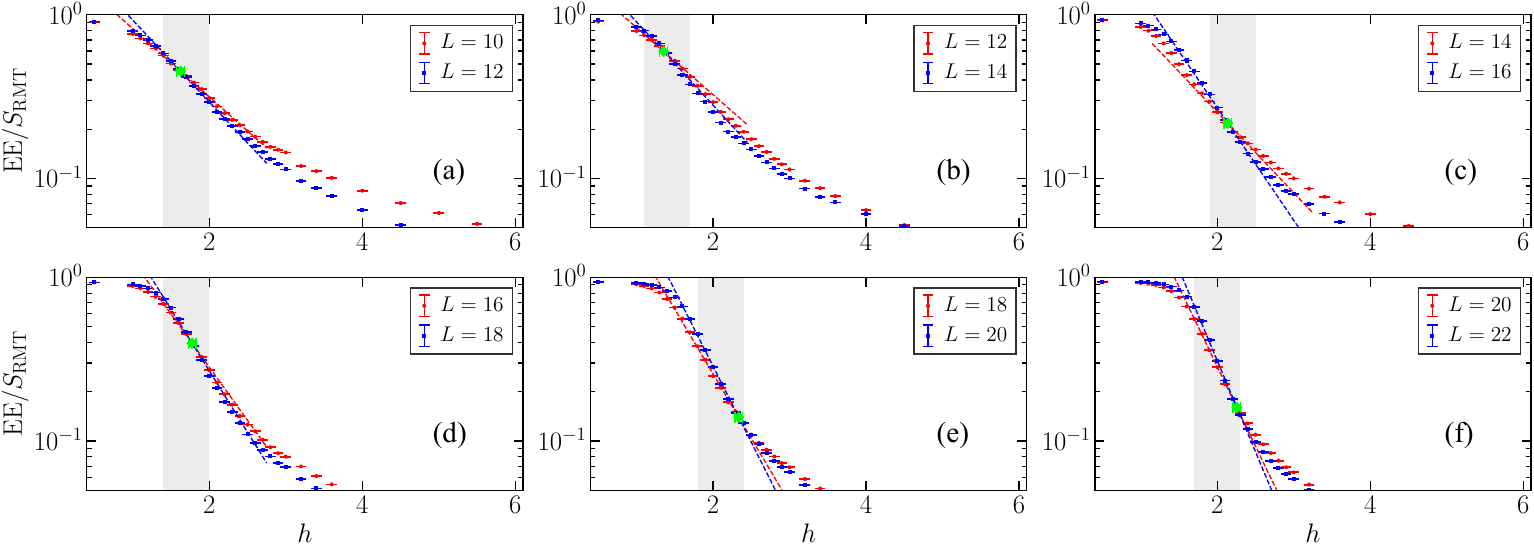}
    \caption{Finite-size scaling analysis of the MBL transition for the quasiperiodic Heisenberg chain using the half-chain entanglement entropy, EE$/S_{\rm RMT}$, for system sizes $L=10, 12, \dots, 22$. The panels show the crossing points $h_{\rm c}(L, L+2)$ for consecutive system sizes. The critical point for each pair of system sizes is determined by fitting curves in the gray region and finding their intersection, which is marked by the green diamond. We observe a non-monotonic drift of these points with increasing system size (see Fig.~\ref{fig:s4}).}
    \label{fig:s3}
\end{figure*}

\textit{\textbf{Participation Entropy (PE).}}
A generalization of the inverse participation ratio to the interacting many-body setting, the participation entropy is defined as~\cite{luca_ergodicity_2013,luitz_many-body_2015,mace_multifractal_2019} (in the Shannon limit)
\begin{equation}
\text{PE} = -\sum_i p_i \ln p_i,
\end{equation}
where $p_i = |\langle i|\psi\rangle|^2$ are the normalized probabilities of the wave function $|\psi\rangle$ in the computational (Fock) basis $\{|i\rangle\}$. A perfectly delocalized state has $\text{PE}= \ln \mathcal{N}$, and an Anderson localized state has $\text{PE} = \rm{const}$. In both phases, $\text{PE} = D_1 \ln{\cal{N}} + b_1$ with $\mathcal{N}$ the Hilbert space dimension. $D_1 =1$ in the ergodic regime with a negative subleading correction $b_1 <0$, whereas $0 < D_1 <1$ and $b_1 >0$ 
in the MBL regime, reflecting multifractality~\cite{mace_multifractal_2019}. The scaling $\text{PE}/\ln{\cal N}$ thus shows a crossing corresponding to the change of sign of the subleading correction.\\

\textit{\textbf{Extreme Magnetization (EM).}} The extreme statistics of the local magnetizations (EM) was recently proposed as a probe of localization~\cite{laflorencie_chain_2020,colbois_breaking_2023}. It focuses on the minimal deviation from perfect polarization, defined as
\begin{equation}
\delta_{\min} = \min_{i} \left(\tfrac{1}{2} - |\langle S_i^z \rangle|\right),
\end{equation}
which identifies the most polarized site in a given eigenstate. Here, we focus on the typical value of $\delta_{\min}$ as an indicator of the ergodic to MBL transition. 
In the thermal phase, ETH predicts that local magnetizations fluctuate around zero, yielding finite $\delta_{\min}^{\rm typ} \rightarrow 1/2$. Due to the particular energy window averaging, here we have $\delta_{\min}^{\rm typ}$ finite in the ergodic phase but a non-Gaussian distribution for $\langle S_i^z \rangle$~\cite{aramthottil_false_2025}. In the localized phase, chain breaks appear in the form of strongly polarized sites, driving $\delta_{\min}^{\rm typ}\!\to\!0$, decaying as a power-law with increasing system sizes.

\section{Finite-size scaling of standard observables}
\label{sec:fss}

We perform a finite-size scaling analysis of the standard observables to estimate the ergodic-MBL transition. For each system size, we identify the crossing points of pairs of curves corresponding to consecutive system sizes, $h_{\rm c}(L,L+2)$. To obtain these crossings robustly, we employ a bootstrap procedure, fit the curves in a small region around the nominal crossing, and compute their intersection.  

\begin{figure}
    \centering
    \includegraphics[width=0.68\columnwidth]{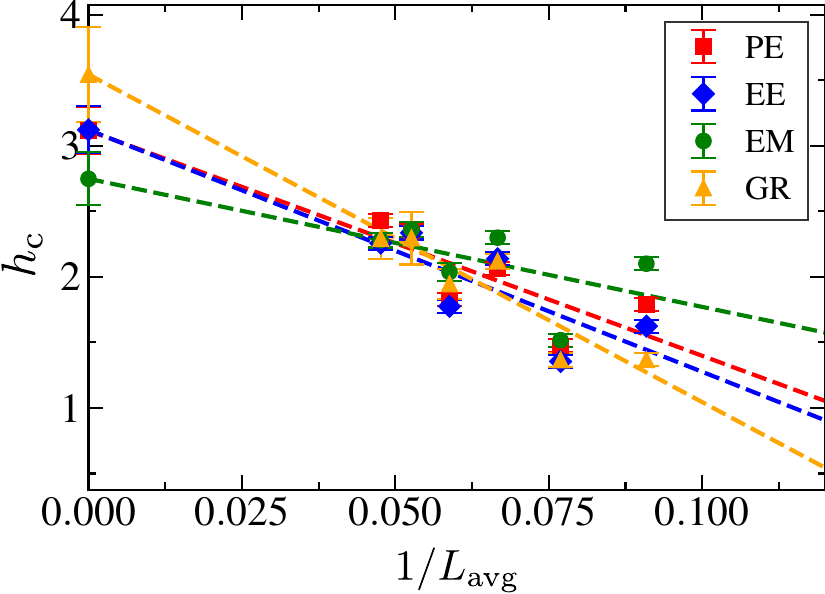}
    \caption{Extrapolation of crossing points for standard observables (PE, EE, EM, GR) for the  quasiperiodic Heisenberg chain. The plot shows the critical field $h_{\rm c}$ {\emph{vs.}} the inverse average system size $1/L_{\rm avg}$. The dashed lines represent the corresponding linear fits.}
    \label{fig:s4}
\end{figure}

\begin{table}
    \centering
    \begin{tabular}{l c}
        \toprule
        Observable & Extrapolated $h_{\rm c}(L_{\rm avg}=\infty)$ \\
        \midrule
        Participation entropy (PE) & $3.12(18)$ \\
        Entanglement entropy (EE) & $3.12(18)$ \\
        Extreme magnetization (EM) & $2.75(20)$ \\
        Gap ratio (GR) & $3.54(36)$ \\
        \bottomrule
    \end{tabular}
    \caption{Estimated critical field strengths $h_{\rm c}$ for the ergodic-MBL transition in the quasiperiodic Heisenberg chain. The values are obtained by extrapolating the crossing points for each observable to the thermodynamic limit using a bootstrap analysis~\cite{SM}.}
    \label{tab:hc_summary}
\end{table}
Figs.~\ref{fig:s3} and \ref{fig:s4} illustrate the finite-size scaling analysis. Fig.~\ref{fig:s3} shows the half-chain EE as a function of $h$ for pairs of system sizes, with the crossing search regions highlighted in gray. Fig.~\ref{fig:s4} plots the resulting crossing points $h_{\rm c}(L,L+2)$ against the inverse average system size, $1/L_{\rm avg}$, for the different observables. Linear fits to the crossing points (dashed lines) fail to converge to a unique value as $1/L_{\rm avg} \to 0$, instead spanning $h_{\rm c} \simeq 2.75$--$3.54$, with an average critical field $h_{\rm c}^{\rm standard} = 3.13(32)$.
 Unlike the random-field XXZ chain, we do not observe a monotonic drift of the crossing points with system size, an effect attributed to the periodic boundary conditions in the quasiperiodic potential. To obtain the critical field in the thermodynamic limit, $h_{\rm c}(L_{\rm avg} = \infty)$, we perform a bootstrap on the set of crossing points for each observable. Tab.~\ref{tab:hc_summary} summarizes the resulting values for PE, EE, EM, and GR.

\vspace{1cm}
\setcounter{section}{0}
\setcounter{secnumdepth}{3}
\setcounter{figure}{0}
\setcounter{equation}{0}
\renewcommand\thesection{S\arabic{section}}
\renewcommand\thefigure{S\arabic{figure}}
\renewcommand\theequation{S\arabic{equation}}

\onecolumngrid

\begin{center}
    \bfseries \large Supplemental Material: Long-range resonances in quasiperiodic many-body localization
\end{center}
This Supplemental Material provides additional analyses and technical details supporting the results presented in the main text. 
In Sec.~\ref{sec:xi_typ_consistency}, we examine the consistency of the extracted typical correlation length $\xi^z_{\rm typ}$ across different fitting windows. 
In Sec.~\ref{sec:field_diffs}, we analyze the distributions of the onsite fields and their differences, and discuss the impact of periodic boundary conditions. 
In Sec.~\ref{sec:non_interacting}, we show that fat-tailed distributions do not arise in the noninteracting limit. 
Finally, in Sec.~\ref{sec:uncertainties}, we describe the procedure used to estimate statistical uncertainties on the various observables.

\section{Consistency of $\xi^z_{\rm typ}$ across fitting windows}
\label{sec:xi_typ_consistency}

\begin{figure}[h!]
    \centering
    \includegraphics[width=\linewidth]{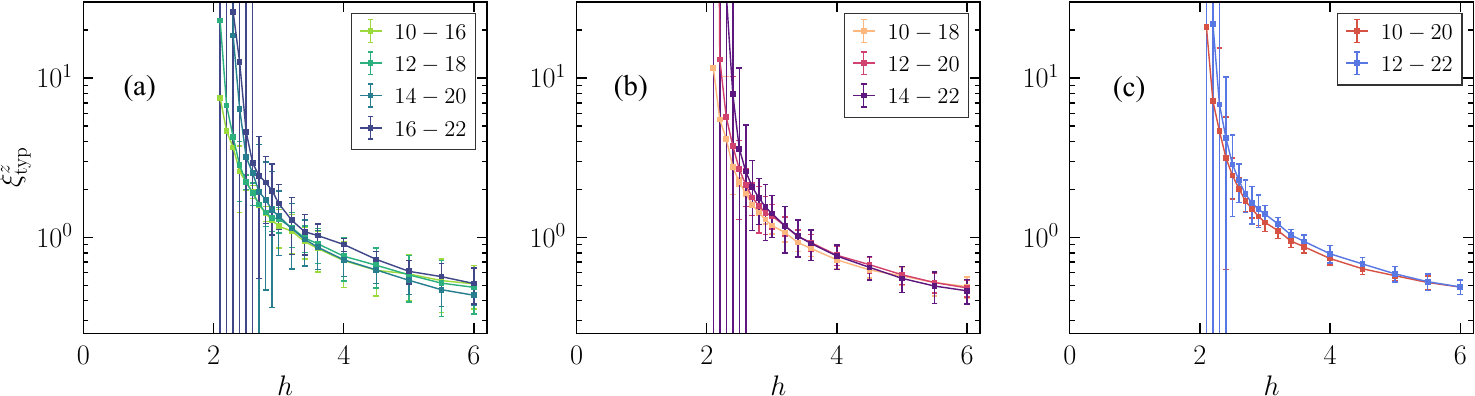}
    \caption{Typical correlation length $\xi^z_{\rm typ}$ as a function of the quasiperiodic field strength $h$
        extracted from (a) 4-point, (b) 5-point, and (c) 6-point sliding-window fits 
        of $\overline{\ln|C^{zz}_{L/2}|}$ for $\Delta = 1$.
        The consistent behavior across different window sizes confirms the robustness 
        of the behavior of $\xi^z_{\rm typ}(L)$, with smaller error bars for larger fitting windows. Though we focus on \emph{typical} values, this supports the presence of a crossover below $h \sim 4-5$.    }
    \label{fig:xi_typ_sliding}
\end{figure}

To assess the robustness of the extracted typical correlation length $\xi^z_{\rm typ}$, 
we perform sliding-window fits of $\overline{\ln|C^{zz}_{L/2}|}$ as a function of system size $L$ 
over several consecutive $L$ intervals. 
Figs.~\ref{fig:xi_typ_sliding}(a--c) show the resulting $\xi^z_{\rm typ}(h)$ obtained from 
4-point, 5-point, and 6-point fits, respectively. 
In all cases, $\xi^z_{\rm typ}$ displays a consistent dependence on the quasiperiodic field strength $h$, 
with only minor variations between the 3 different sizes of the fitting windows. 
The error bars systematically decrease with increasing window size, 
reflecting the improved statistical stability of longer fitting intervals. 
Importantly, $\xi^z_{\rm typ}$ grows with system size when approaching $h \sim 4-5$ from the MBL side, 
signaling the onset of instability in the localized phase.

\section{Quasiperiodic potential: distributions and boundary conditions}
\label{sec:field_diffs}

\begin{figure}[htbp]
    \centering
    \includegraphics[width=.7\columnwidth]{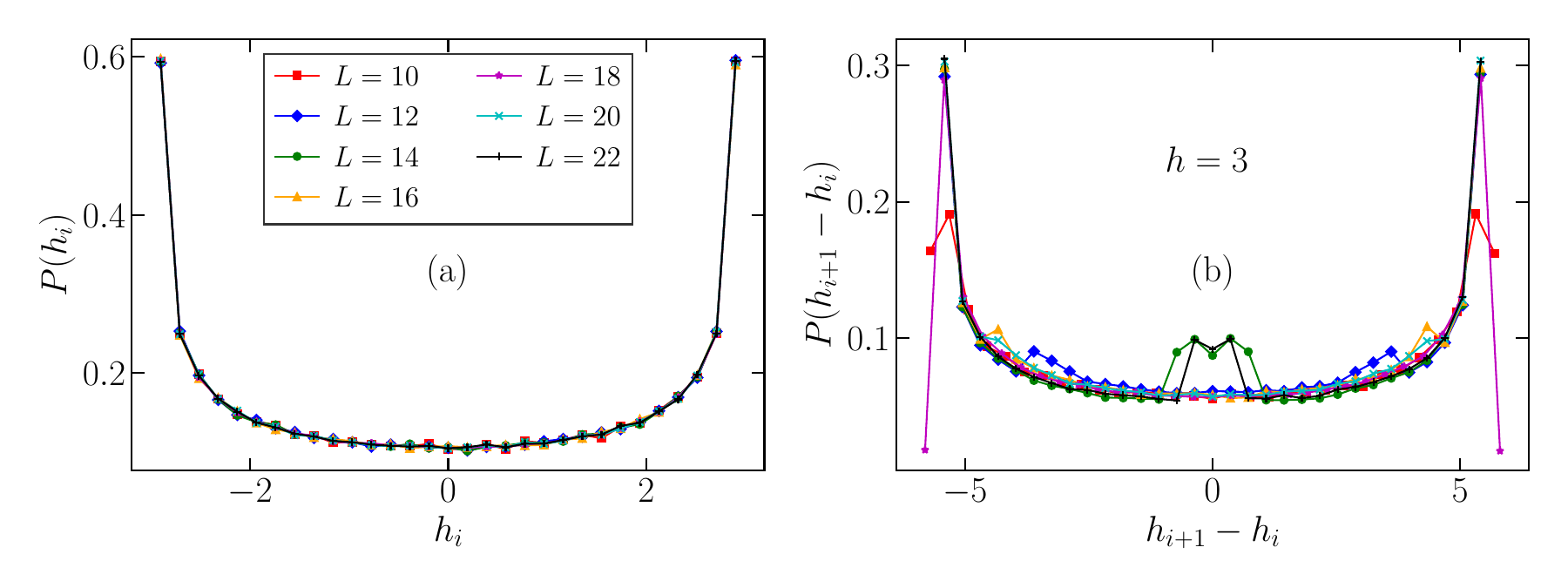}
    \caption{\label{fig:fields_dist} (a) Probability distribution $P(h_i)$ of the quasiperiodic field values for system sizes $L=10\!-\!22$ at $h=3$, obtained from multiple realizations of the random phase $\phi\in [0,\,2\pi)$. The distribution is bounded within $[-h,h]$ and exhibits peaks at both extrema, as expected from Eq.~\eqref{eq:arcsine}. (b) Probability distribution $P(\delta_i=h_{i+1}-h_i)$ of the difference between adjacent quasiperiodic field values at $h=3$, also collected over different phase samples. $P(\delta_i)$ also follows a similar arcsine law, but with a slightly different width, given by 
$2h\sin\left[\pi\beta\right]\approx 1.864h$. The ``anomalous'' feature that appear only for $L=14$ and $L=22$ comes from the PBC which yields a softer bond at the boundary $2h\sin\left[\pi\beta(L-1)\right]$, see text and Tab.~\ref{tab:deltaL}.}
\end{figure}

Fig.~\ref{fig:fields_dist}(a) shows the arcsine distribution followed by onsite fields, 
\begin{equation}
P(h_i) = \frac{1}{\pi\sqrt{h^2 - h_i^2}}, \qquad |h_i|\le h,\label{eq:arcsine}
\end{equation}
confined within $[-h,h]$ and exhibiting characteristic peaks at the extrema $h_i=\pm h$.  
For neighboring sites, one can define the bond variable $\delta_i$, for $i<L$ as
     \begin{eqnarray}
         \delta_i&=&h_{i+1}-h_i\nonumber\\
         &=&h\left(\cos\left[2\pi\beta(i+1)+\phi\right]-\cos\left[2\pi\beta i+\phi\right]\right)\nonumber\\
         &=&-2h\sin\left[\pi\beta\right]\sin\left[\pi\beta(2i+1)+\phi\right].\label{eq:deltai}
     \end{eqnarray}
     At the boundary of a closed ring (PBC), the $(L+1)^{\rm th}$ site is identified with the first one, and therefore one has at the boundary 
     \begin{eqnarray}
         \delta_L&=&h_{1}-h_L\nonumber\\
         &=&h\left\{\cos\left[2\pi\beta+\phi\right]-\cos\left[2\pi\beta L+\phi\right]\right\}\nonumber\\
         &=&2h\sin\left[\pi\beta(L-1)\right]\sin\left[\pi\beta(L+1)+\phi\right].\label{eq:deltaL}
     \end{eqnarray}
     While the terms $\sin\left[\pi\beta(2i+1)+\phi\right]$ in Eq.~\eqref{eq:deltai} and $\sin\left[\pi\beta(L+1)+\phi\right]$ in Eq.~\eqref{eq:deltaL}  are  distributed according to the arcsine law Eq.~\eqref{eq:arcsine} with $h=1$, the prefactors differ. It is $2h\sin\left[\pi\beta\right]\approx 1.864h$ for the $L-1$ ``bulk'' bonds, while it is $2h\sin\left[\pi\beta(L-1)\right]$ for the ``boundary'' link. Below in Tab.~\ref{tab:deltaL} we provide the numerical values for the  $\delta_L$
prefactor, which is strongly suppressed when $L-1$ coincides with a Fibonacci number ($1, \,2, \,3, \,5, \,8, \,13,\, 21, \,34, \,55,\, 89,\, 144,\,\ldots$), marked in red for $L=14,\,22$.
     \begin{table}[h!]
\centering
\begin{tabular}{cc}
\toprule
$L$ & $\left|\sin\!\left[\pi\beta(L-1)\right]\right|$ \\
\midrule
10 & 0.98090 \\
12 & 0.59191 \\
{\color{red}{14}} & {\color{red}{0.10799}} \\
16 & 0.75117 \\
18 & 0.99979 \\
20 & 0.72325 \\
{\color{red}{22}} & {\color{red}{0.06682}} \\
\bottomrule
\end{tabular}
\caption{Prefactor $\left|\sin\!\left[\pi\beta(L-1)\right]\right|$ of the boundary term that appear in $\delta_L$, Eq.~\eqref{eq:deltaL} for the 7 different system sizes considered. In red we shown the two cases when $L-1$ coincides with a number of the Fibonacci sequence, for which the boundary term is strongly suppressed.\label{tab:deltaL}}
\end{table}

The distribution of field differences is only slightly affected by the effect of PBC, only on the boundary link for some special sizes. This is however an $1/L$ effect which is expected to disappear at large system sizes.

\section{Absence of Fat Tails in the Noninteracting Limit}
\label{sec:non_interacting}
In order to show that the fat-tail shape found for the correlations is a direct consequence of finite interactions, we compare  the interacting quasiperiodic Heisenberg case to its noninteracting XX limit (Eq.~(1) in the main text with $\Delta=0$), which shows no evidence of fat tails in the correlation data. This confirms that the observed cat states and the associated instability are emergent phenomena induced by interactions. The data are obtained from free-fermion calculations covering the full many-body spectrum for each value of $\phi$.

\begin{figure}[htbp]
    \centering
    \includegraphics[width=\columnwidth]{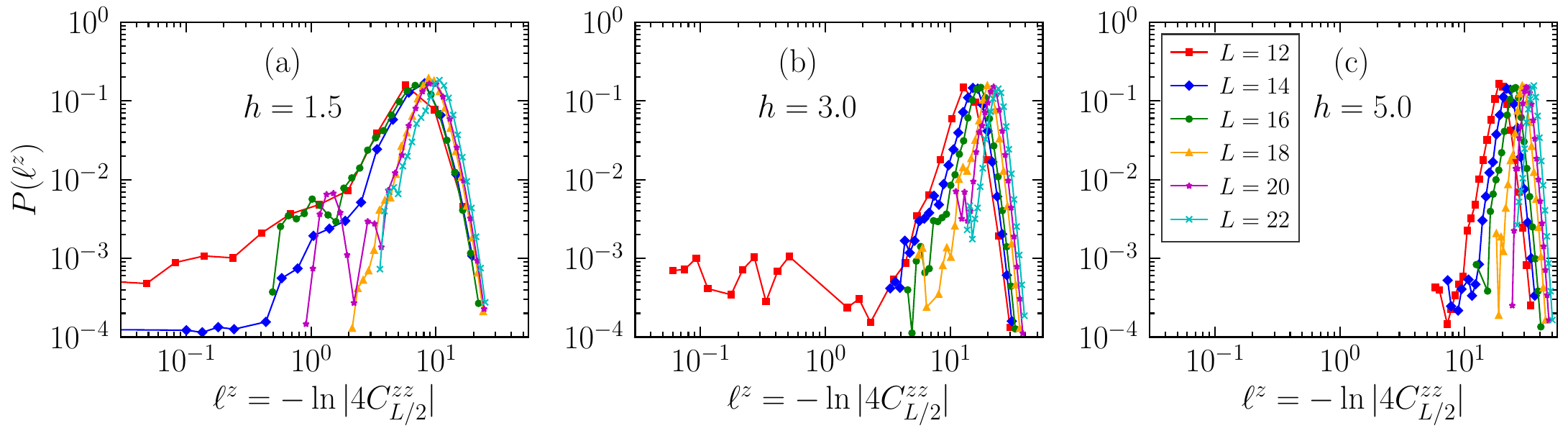}
    \caption{\label{fig:noninteracting}
    Probability distribution $P(\ell^z)$ of the rescaled longitudinal correlation $\ell^z = -\ln|4C_{L/2}^{zz}|$ for the noninteracting quasiperiodic XX chain (Eq.~(1) in the main text with $\Delta=0$). 
    Data are shown for system sizes $L=12,14,16,18,20,22$ and field strengths $h=1.5,3.0,5.0$, all within the localized regime. 
    The distributions exhibit a fast decay (with increasing system sizes) at large $\ell^z$ , indicating the absence of fat tails and of rare, large correlations.
    }
\end{figure}

\begin{figure}[h!]
    \centering
    \includegraphics[width=0.5\linewidth]{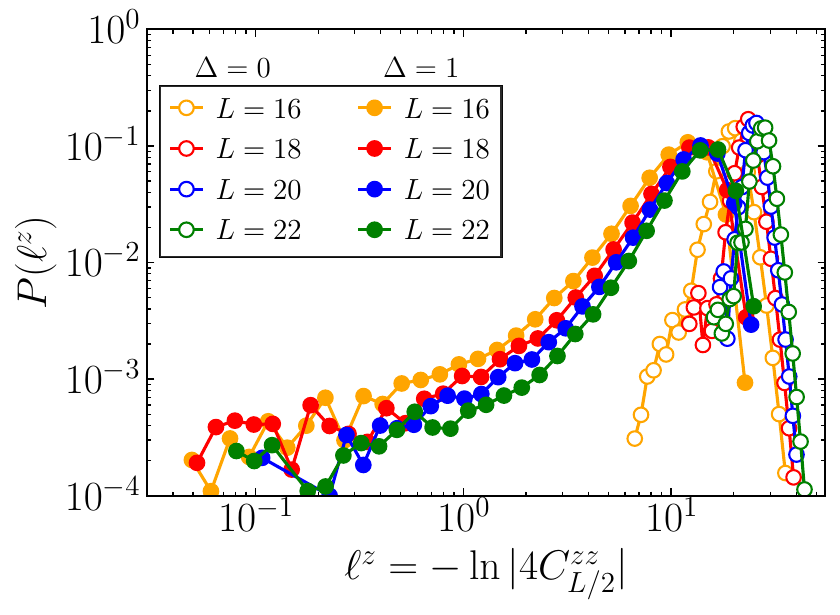}
    \caption{Probability distribution $P(\ell^z)$ of $\ell^z = -\ln|4 C^{zz}_{L/2}|$ for system sizes $L = 16, 18, 20, 22$ at field $h=3.6$. The fat tails at small $\ell^z$ (large correlations) in the interacting case ($\Delta=1$, $\epsilon=0.5$) are highlighted by the contrast with the noninteracting case ($\Delta=0$, full spectrum), which lacks such tails.}
    \label{fig:histogram_comparison}
\end{figure}

Fig.~\ref{fig:noninteracting} shows the probability distribution $P(\ell^z)$ of the rescaled longitudinal correlation $\ell^z = -\ln|4C_{L/2}^{zz}|$ for the noninteracting chain at various system sizes $L$ and field strengths $h$. 
We focus on the localized phase, considering $h=1.5$, $3.0$, and $5.0$, all above the single-particle localization threshold $h=1$ of the Aubry–André model. 
For all parameters shown, the distributions exhibit a clear exponential decay at large $\ell^z$, demonstrating that large correlations are exponentially suppressed. 
This behavior is characteristic of conventional Anderson localization, where correlations decay exponentially with distance and rare long-range events are absent (see also Ref.~\cite{colbois_statistics_2024}). 
The lack of fat tails in these distributions confirms that no rare, system-spanning correlations occur in the noninteracting localized regime, in stark contrast with finite $\Delta$. We thus have no long-distance cat states at $\Delta=0$, differing from the interacting case discussed in the main text. This is further highlighted by Fig.~\ref{fig:histogram_comparison}, which shows a comparison at $h = 3.6$ of the distribution of correlators at $\Delta = 0$ and $\Delta = 1$.

\section{Uncertainties in Observables}
\label{sec:uncertainties}

To estimate uncertainties in our observables, we combine averages over eigenstates, spatial indices, and quasiperiodic phase realizations. 
For a generic observable $\mathcal{O}$, we first define the average over eigenstates and spatial indices (cuts for entanglement entropy or site pairs for correlations) for each sample $s$ as
\begin{equation}
\overline{\mathcal{O}}_s = \frac{1}{N_{\mathrm{eig}} N_{\mathrm{sites}}} \sum_{e=1}^{N_{\mathrm{eig}}} \sum_{\alpha=1}^{N_{\mathrm{sites}}} \mathcal{O}_{e,\alpha}^{(s)},
\end{equation}
where $N_{\mathrm{sites}} = L/2$ for both entanglement entropy and correlation measurements, and $s=1,\dots,N_{\mathrm{samp}}$ labels the phase samples.

The standard deviation of $\overline{\mathcal{O}}_s$ across samples quantifies sample-to-sample fluctuations:
\begin{equation}
\sigma_{\mathcal{O}} = \sqrt{\frac{1}{N_{\mathrm{samp}}-1} \sum_{s=1}^{N_{\mathrm{samp}}} \left( \overline{\mathcal{O}}_s - \overline{\mathcal{O}} \right)^2 },
\end{equation}
where the sample-averaged observable is
\begin{equation}
\overline{\mathcal{O}} = \frac{1}{N_{\mathrm{samp}}} \sum_{s=1}^{N_{\mathrm{samp}}} \overline{\mathcal{O}}_s.
\end{equation}

Finally, the uncertainty (error bar) reported for $\mathcal{O}$ is the standard error of the mean,
\begin{equation}
\Delta \mathcal{O} = \frac{\sigma_{\mathcal{O}}}{\sqrt{N_{\mathrm{samp}}}},
\end{equation}
which accounts for fluctuations from both eigenstate variations and sample-to-sample variations.

For the KL divergence of the GR distribution (see End Matter), the main challenge is to estimate errors reliably due to finite sampling and histogram effects. 
We compute $30$ independent KL values for each parameter set, randomly sampling $20$k ratios for $L<12$ and $30$k ratios otherwise. 
Two different histogram bin numbers are considered to account for binning effects. 
The standard deviation of these $30$ KL values is reported as the error bar, providing a robust estimate of the uncertainty. 
Numerically, empty bins are correctly assigned zero weight, ensuring stability even in sparse regions.

Uncertainties in all fitting procedures are quantified using a parametric bootstrap. In this approach, $N_{\mathrm{boot}}=1000$ synthetic datasets are generated by resampling within the estimated errors, and each is refitted independently. The resulting distribution of parameters captures statistical fluctuations and finite-size variability; we report its median as central value and the 95\% confidence interval as uncertainty. This procedure is applied uniformly to all extrapolations and fits.

\end{document}